\documentclass[twocolumn,superscriptaddress,nofootinbib]{revtex4-2}
\usepackage{amsmath}
\usepackage{mathtools}
\usepackage{amssymb}
\usepackage[utf8]{inputenc}
\usepackage{physics}
\usepackage{tabularx}
\usepackage{overpic}
\usepackage{balance}
\usepackage{color}
\usepackage{xcolor}
\usepackage{graphicx}
\usepackage{import}
\usepackage{tabularx}
\usepackage{dcolumn} 
\usepackage{bm}      
\usepackage{array}   
\usepackage{amsfonts}
\usepackage{tikz}
\usepackage[T1]{fontenc}
\usepackage{float}
\usepackage{tabularray}[2021-06-01]

\usepackage[english]{babel}
\usepackage[bookmarks=false,colorlinks,citecolor=red]{hyperref}

\usepackage[utf8]{inputenc}
\usepackage[T1]{fontenc}
\usepackage{mathptmx}

\usepackage[normalem]{ulem}

\include{MyCommand}

\DeclareUnicodeCharacter{2212}{-}

\makeatletter
\newcommand*{\addFileDependency}[1]{
  \typeout{(#1)}
  \@addtofilelist{#1}
  \IfFileExists{#1}{}{\typeout{No file #1.}}
}
\makeatother

\include{MyCommand}

\begin{document} 


\title{Dynamical quantum phase transitions from relative negativity of Wigner function
} 




\author{Daniel ~J.~ Nader}
\email{daniel.juliannader@upol.cz}
 \affiliation{Department of Optics, Faculty of Science, Palack\'y University, Olomouc 77900, Czech Republic}

\author{Jakub Novotn\'y}
\email{jakub.novotny@matfyz.cuni.cz}
\affiliation{
 Institute of Particle and Nuclear Physics, Faculty of Mathematics and Physics,
Charles University, V Hole\v{s}ovi\v{c}k\'ach 2, 18000 Prague, Czech Republic}


\author{Pavel Str\'ansk\'y}
\email{pavel.stransky@matfyz.cuni.cz}
 \affiliation{
 Institute of Particle and Nuclear Physics, Faculty of Mathematics and Physics,
Charles University, V Hole\v{s}ovi\v{c}k\'ach 2, 18000 Prague, Czech Republic}

\author{Pavel Cejnar}
\email{pavel.cejnar@matfyz.cuni.cz}
\affiliation{
 Institute of Particle and Nuclear Physics, Faculty of Mathematics and Physics,
Charles University, V Hole\v{s}ovi\v{c}k\'ach 2, 18000 Prague, Czech Republic}
  \author{Radim Filip}
   \email{filip@optics.upol.cz} 
\affiliation{Department of Optics, Faculty of Science, Palack\'y University, Olomouc 77900, Czech Republic}
 

\begin{abstract}
Dynamical quantum phase transitions (DQPTs), manifested as kinks in the dynamical rate function, have emerged as a central topic in nonequilibrium quantum physics. In ideally isolated spin–boson systems, DQPTs have been predicted in the feasible quantum Rabi model (QRM), yet their experimental observation in realistic open-system dynamics and their understanding connected to quantum non-Gaussian aspects in bosonic systems remain challenging as they require access to the full quantum state of the spin–boson system over time. Here, we analyze dynamical kinks in  a rate function of the bosonic part expressed in terms of the directly measurable bosonic overlap. 
Analyzing these kinks from a phase space perspective, we decompose the bosonic rate function to derive the relative Wigner negativity, a quantum non-Gaussianity quantifier that inherits the characteristic kinks, thereby uncovering their genuine quantum origin. 
These kinks in the relative Wigner negativity  exhibit the corresponding scaling with system size and are not critically sensitive to weak thermal occupation and loss. 
\end{abstract}


\maketitle

Quantum phase transitions (QPTs) and critical phenomena are central to the emergence of novel effects in quantum systems \cite{Sachdev_2011,carr2010understanding}. At the unphysical zero temperature and idealized infinite-size limit, QPTs are defined as nonanalytic points in the ground state energy as a function of a control parameter and manifest as qualitative changes in the pure ground state \cite{Sachdev_2011,carr2010understanding,vojta2003quantum,PhysRevE.74.031123,CEJNAR2009210,trenkwalder2016quantum}. Critical behavior can also extend into excited states, giving rise to excited-state quantum phase transitions (ESQPTs), characterized by nonanalyticities in the density of states \cite{cejnar2021excited,SantosESQPT,wang2021esqpt,cejnar2006monodromy,caprio2008excited,stransky2014excited,stransky2015excited,Macek2019}. Both ground state QPTs and ESQPTs are equilibrium phenomena associated with stationary states of idealized, time-independent closed systems. Although the theory of equilibrium QPTs is vast, preparing, detecting and characterizing the idealized stationary states required to verify critical behavior remain challenging, as this demands idealized infinite size and extremely low temperatures that leads to long relaxation times near criticality, in competition with loss and noise. In contrast, experimental studies rely on directly measurable witnesses to observe precursors of equilibrium criticality at realistic finite-size systems~\cite{cai2021observation,ZhengPRL}. In addition, these witnesses may advantageously provide information about nontrivial nonclassical aspects behind the criticality. In spin–boson systems, the negativity of the Wigner function \cite{hillery1984distribution} is a directly measurable witness based on displaced parity detection \cite{bertet2002direct,wang2009decoherence,wang2016schrodinger} and proves nontrivial Wigner function structure. The Wigner function negativity was already partially analyzed in spin-boson systems \cite{catrabi,LopezPena2021} exhibiting criticality.

Dynamical quantum phase transitions (DQPTs) \cite{Heyl2013}, in contrast, are nonequilibrium critical phenomena that manifest as kinks in the dynamical rate function, indicating nonanalytic behavior in the real-time evolution \cite{Heyl2017DynamicalQP,xu2024quantum}. They typically arise during the transient dynamics following sharp quantum quenches  \cite{Vajna2014,Zvyagin2016DynamicalQP}. DQPTs have been theoretically predicted in spin chains \cite{vosk2014dynamical,divakaran2016tuning,sen2005dynamical,PhysRevB.96.104436}, as well as in discrete \cite{YU2026170344,lohr2025dynamical,complextime} and continuous-variable settings \cite{Puebla2020}. Beyond theoretical predictions, DQPTs can be more accessible experimentally, as they are tied to nonequilibrium protocols, circumventing the preparation of idealized stationary states required for equilibrium QPTs. Furthermore, the characteristic nonanalyticity in the dynamical rate function can emerge already at small system sizes \cite{Heyl2017DynamicalQP,zhou2021exact,novotny2026tracing,Bojan2018}, contrary to equilibrium QPTs, whose characteristic nonanalyticity in the ground state energy is expected only in the idealized infinite-size limit.   Indications of sharp kinks in dynamical rate functions have been already observed experimentally using discrete-variable quantum simulators of spin systems  \cite{DQPTtrappedions,Chen2020,Guo2019,xu2020probing}. For small number of pseudo-spins, tomography of mixed quantum states \cite{Guo2019} or positive operator-valued measures \cite{DQPTtrappedions} allow the survival probability (or Loschmidt echo) to be measured, enabling the corresponding dynamical rate function to be reconstructed.

In spin–boson systems, however, the situation is markedly different, as the standard diagnosis requires full state reconstruction of the entire infinite-dimensional spin–boson system over time, which is significantly more challenging and hinders the direct observation of DQPTs. Furthermore, the kinks in the dynamical rate functions can be fragile under mixed-state dynamics \cite{bhattacharya2017mixed,lang2018dynamical,Mera2018,Sedlmayr2018}, which naturally arise in any experimental conditions and conclusively confirming the nonanalyticity of the characteristic kinks of DQPTs remains challenging for mixed-state dynamics \cite{Liu2024}.  A natural approach to circumvent these difficulties is to
investigate kinks only in the reduced subsystems
\cite{karch2025probing}. Recently, kinks were observed experimentally in the reduced spin rate function \cite{bian2026temporal}.
 On the other hand, the bosonic sector offers a complementary
phase-space perspective, as  bosonic Wigner functions can be directly
measured without full state tomography
\cite{bertet2002direct,wang2009decoherence,wang2016schrodinger}. However, a direct understanding of DQPT manifestations in realistic conditions by quantum non-Gaussian aspects of the bosonic Wigner function remains largely underexplored \cite{photonics10111184}.

The Jaynes–Cummings model \cite{jaynes1963comparison,Shore01071993,larson2021jaynes} provides a minimal experimental framework for spin-boson interactions, but it is not well suited for the exploration of DQPTs. Although it exhibits a first-order equilibrium QPT associated with level crossings \cite{liu2024quantum}, its dynamical response is essentially insensitive to the choice of quench parameters (see Appendix \ref{Fisher}). In contrast, the quantum Rabi model (QRM) \cite{rabi1936process} exhibits a second-order quantum phase transition \cite{cai2021observation,PueblaQPT_QRM}, characterized by universal dynamical scaling \cite{polkovnikov2005universal}. This makes the QRM the simplest platform for the exploration of DQPTs in spin-boson systems. The QRM is currently available in a variety of experimental platforms, including superconducting circuits, photonic systems, semiconductors, and trapped ions \cite{forn2010observation,braumuller2017analog,yoshihara2017superconducting,langford2017experimentally,crespi2012photonic,todorov2009strong,gunter2009sub,lv2018quantum,cai2021observation}. While DQPTs have already been predicted in the QRM \cite{Puebla2020}, their experimental observation and deeper understanding based on quantum non-Gaussian aspects of bosonic Wigner function remain challenging, as the current formulation involves the entire spin-boson system.

In this work, we derive a Wigner-negativity quantifier that explains the DQPTs of the QRM by quantum non-Gaussian effects in phase space. First, we verify nonequilibrium criticality in the ideal QRM from pure states by the analysis of Fisher zeros and their scaling. We also demonstrate that DQPTs in the QRM already manifest in the directly measurable bosonic rate function, while the off-resonant ancillary qubit that induces the nonlinear bosonic dynamics is traced out. Second, we derive the relative Wigner negativity directly from the decomposition of the bosonic rate function, providing a quantifier of quantum non-Gaussian effects for the phase space analysis of the kinks in the bosonic rate function. Third, we analyze the QRM dynamics under experimentally relevant bosonic loss. Fourth, importantly, we show that the sharpness of the kink in the relative Wigner negativity exhibits the same scaling with bosonic loss as the kink in the bosonic rate function, confirming that the nonequilibrium criticality in the QRM is a genuine quantum non-Gaussian phenomenon that disappears as the Wigner negativity is suppressed.

 Let us consider the Hamiltonian of the QRM \cite{rabi1936process,catrabi,Pan_2010,Braak2011}
\begin{equation}
\label{HQRM}
\hat{H}= \hat{a}^\dagger\hat{a} + \frac{R}{2}\hat{\sigma}_z
+ \frac{\lambda\sqrt{R}}{2}\,(\hat{a}^\dagger + \hat{a})\hat{\sigma}_x ,
\end{equation}
where $\hat{a}^\dagger$ and $\hat{a}$ are the bosonic creation and annihilation operators, and $\hat{\sigma}_i$ denote the Pauli matrices acting on the qubit subspace. The bosonic frequency is fixed to unity, $R$ corresponds to the qubit frequency and $\lambda$ is the coupling strength.  The Hamiltonian conserves a
generalized parity $\hat{\Pi}=(-1)^{\hat{n}}\hat{\sigma}_z$ where $\hat{n}=\hat{a}^\dagger\hat{a}$ is the number of bosons.

To obtain an approximation analogous to the infinite-size limit, where the QPTs are well defined, it is possible to derive an effective semiclassical Hamiltonian \cite{Stransky2021,catdeath2024,irish2014oscillator,Puebla2016}.
For this purpose, we can take the expectation value of (\ref{HQRM}) in the oscillator subspace $\hat{H}_\alpha=\langle\alpha|\hat{H}|\alpha\rangle$, where $|\alpha\rangle=\hat{D}(\alpha)|0\rangle={\rm exp}(\alpha\hat{a}^\dagger-\alpha^*\hat{a})|0\rangle$ is a coherent state with displacement parameter $\alpha=\sqrt{\frac{R}{2}}(x+ip)$. The eigenvalues of the resulting expectation value correspond to the effective semiclassical Hamiltonians of the spin-up and spin-down subspaces. We can write these effective semiclassical nonlinear Hamiltonians in terms of the real and imaginary parts of the displacement parameter as $x$ and $p$, obtaining
\begin{equation}
\label{Hsemiclassical}
h_{eff}^{\pm}=\frac{\langle H \rangle}{R} =\frac{p^2+x^2}{2}\pm \frac{1}{2}\sqrt{1 + 2\lambda^2x^2}\,.
\end{equation}

Since we are interested in the low-energy spin subspace, we choose the negative branch of the square root. Given the definition of the canonical variables $x$ and $p$ in terms of the displacement parameter $\alpha$, the correspondence between the quantum Hamiltonian (\ref{HQRM}) and the effective Hamiltonian (\ref{Hsemiclassical}) improves as the qubit frequency $R$ increases \cite{catdeath2024,Stransky2021}. Here, the qubit frequency $R$ plays the role of the system size. In the limit $R\to\infty$, the qubit in (\ref{HQRM}) plays the role of an off-resonant ancilla that introduces the nonlinear effects in the bosonic system. The potential in the semiclassical Hamiltonian (\ref{Hsemiclassical}) clearly displays a competition between the quadratic free-oscillator terms and the nonlinear contribution with the square root, which arises from the off-resonant interaction with the qubit. For $\lambda<1$, the quadratic contribution dominates near the origin and the potential has a single minimum at $x=0$. For $\lambda>1$, the curvature at $x=0$ becomes negative and the potential develops a double-well structure. This transition at $\lambda_c=1$, driven by the nonlinear effect induced by an off-resonant qubit, gives rise to the equilibrium second-order QPT of the QRM \cite{PueblaQPT_QRM}, separating the normal and superradiant phases.
The quantization of the Hamiltonian (\ref{Hsemiclassical}) can be considered as a target bosonic nonlinearity that (\ref{HQRM}) tries to emulate. Their correspondence is analyzed in Appendix \ref{correspondence}.

The standard protocol for generating DQPTs \cite{Heyl2017DynamicalQP} is based on quantum quenches. In the context of the QRM, the control parameter that is suddenly varied is the coupling strength $\lambda$. Ideally, the system is prepared in the ground state $|\psi_0\rangle = |E_0(\lambda_0)\rangle$ of the Hamiltonian $\hat{H}(\lambda_0)$, which satisfies $\hat{H}(\lambda_0)|E_0(\lambda_0)\rangle = E_0(\lambda_0)|E_0(\lambda_0)\rangle$, corresponding to an initial coupling $\lambda_0$. At time $t=0$, the coupling strength $\lambda_0$ is abruptly changed to a new value $\lambda_f$, such that the Hamiltonian becomes 
$\hat{H}(\lambda_f)$.
As a consequence of this sudden quench, the system undergoes unitary time evolution, which can be formally written as $|\psi(t)\rangle=\sum_k \langle E_k(\lambda_f)|\psi_0\rangle e^{-iE_k(\lambda_f)t}|E_k(\lambda_f)\rangle$, where $|E_k(\lambda_f)\rangle$ are the eigenstates of the final Hamiltonian satisfying $\hat{H}(\lambda_f)|E_k(\lambda_f)\rangle = E_k(\lambda_f)|E_k(\lambda_f)\rangle$.

During the subsequent idealized unitary dynamics, the traditional quantity for the study of DQPTs is the rate function \cite{Heyl2017DynamicalQP}
\begin{equation}
\label{ratefunction}
\Lambda(t)=-\frac{1}{R}\log |{\mathcal Z}(t)|^2 ,
\end{equation}
where ${\mathcal Z}(t)=\langle\psi_0|\psi(t)\rangle=\sum_k|\langle\psi_0|E_k(\lambda_f)\rangle|^2e^{-iE_k(\lambda_f)t}$ is the survival amplitude (also called Loschmidt amplitude) and $|{\mathcal Z}(t)|^2$ is the survival probability \cite{lerma2018analytical,vanicek2003survival}, defined as the overlap between the initial and the entire spin-boson evolved state. In the idealistic case, DQPTs are identified as kinks in the rate function \cite{Heyl2017DynamicalQP}, i.e., nonanalytic points at which the first derivative $\Lambda'(t)$ becomes discontinuous. To theoretically confirm these nonanalyticities in ideal unitary dynamics, it is essential to extend the domain of time to the complex plane \cite{Peota2021,complextime} by defining $t' = t + i\gamma$. When the zeros of the survival amplitude, $\mathcal{Z}(t')=0$, known as Fisher zeros \cite{Peota2021,fisher1967theory}, approach the real-time axis, they generate kinks in the rate function, confirming the presence of a nonanalyticity i.e. the nonequilibrium criticality \cite{grossmann1969phase,derrida1991zeroes,borrmann2000classification}. 
Unlike the finite-size precursors of equilibrium criticality, the finite-size manifestations of nonequilibrium criticality do not, in general, sharpen gradually with increasing system size.  This is because the survival probability $|{\mathcal Z}(t)|^2$ can vanish already at a given finite system size~\cite{Heyl2017DynamicalQP,zhou2021exact,novotny2026tracing} generating a genuine nonanalytic point in the dynamical rate function, in contrast to the partition function which remains strictly positive at finite sizes~\cite{YangLee}. Both the kinks in the rate function (\ref{ratefunction}) and the Fisher zeros $\mathcal{Z}(t')=0$ provide the formally conclusive indicators of nonequilibrium criticality for finite size and under pure state dynamics.  For mixed-state dynamics, different generalization of the survival probability
$|{\mathcal Z}(t)|^2$ exist  \cite{bhattacharya2017mixed,lang2018dynamical,lang2018dynamical,Mera2018,Sedlmayr2018,Link2020}. However, their analytic continuation to complex-time, necessary for the Fisher zero analysis and the confirmation of nonanalytic behavior, remains numerically challenging \cite{Liu2024}.

In their conventional formulation, the DQPTs of the QRM \cite{Puebla2020} rely on the survival amplitude of the entire pure state of the full spin–boson system, complicating the experimental observation in real open systems with mixed states. Moreover, they do not provide further insight into the underlying phase space structure of the bosonic mode, which is also experimentally accessible. Additionally, due to nonlinearity, the bosonic states might be highly complex and extensive, such that a direct measurement of their critical behavior without full-state tomography will be highly appreciated. This motivates (i)  to investigate kinks purely in the bosonic mode, (ii) the use of directly measurable quantities and (iii) extension to realistic mixed-state dynamics. For only point (i), such a reduction naturally arises by choosing suitable initial product states of the form $|n\rangle \otimes |\downarrow\rangle$, given by the tensor product of Fock states $|n\rangle$ and the spin-down state $|\downarrow\rangle$. For simplicity, we can focus on the vacuum initial state $|\psi_0\rangle = |0\rangle \otimes |\downarrow\rangle$, as it is routinely well approached in the main experimental platforms \cite{lv2018quantum,braumuller2017analog}. For ideal unitary dynamics, due to the well-defined parity $\hat{\Pi}$, the entire spin-boson survival probability reduces exactly to the overlap of the bosonic reduced density matrices,
$|{\mathcal Z}(t)|^2 = |{\mathcal Z}_b(t)|^2 = {\rm Tr}\big[\hat{\rho}_b(0)\hat{\rho}_b(t)\big]=\langle0|\hat{\rho}_b(t)|0\rangle$,
where $\hat{\rho}_b = {\rm Tr}_q(\hat{\rho})$ denotes the partial trace of the full density matrix $\hat{\rho}$ over the qubit degrees of freedom (see Appendix \ref{proof} for a detailed derivation). Therefore, DQPTs rely entirely on the bosonic sector under unitary dynamics, despite the presence of the ancillary off-resonant qubit degree of freedom.  Point (ii) is fulfilled because the bosonic survival probability $|{\mathcal Z}_b(t)|^2={\rm Tr}\big[\hat{\rho}_b(0)\hat{\rho}_b(t)\big]$, expressed now as an overlap integral of realistic and general mixed bosonic states, can be directly measured interferometrically using, for example, a beam-splitter scheme and available parity detection~\cite{filip2002overlap,Gao2018,Gan2020}. Furthermore, the bosonic survival probability expressed as the vacuum occupation probability $|{\mathcal Z}_b(t)|^2=\langle0|\hat{\rho}_b(t)|0\rangle$ can be inferred from the qubit population dynamics induced by a blue-sideband transition \cite{cai2021observation}. Therefore, after thoroughly checking the nonequilibrium criticality confirmed by Fisher zeros for unitary dynamics (see Appendix \ref{Fisher}) we concentrate on the realistic and measurable bosonic survival probability $|{\mathcal Z}_b(t)|^2$ for general mixed states and realistic non-unitary dynamics for point (iii) and its corresponding bosonic rate function $\Lambda_b(t)=-(1/R)\log|{\mathcal Z}_b(t)|^2$.

Moreover, to gain phase space insight beyond the bosonic rate function, we employ the bosonic Wigner function \cite{hillery1984distribution,wignerfunctions}, as it can also be directly measured without full quantum tomography~\cite{bertet2002direct,wang2009decoherence,wang2016schrodinger}.
 The bosonic Wigner function is defined as
\begin{equation}
\label{Wignerqpt}
W(x,p,t) = \frac{R}{\pi} \int_{-\infty}^{\infty} dy\, e^{2ipy R}\, \langle x - y | \hat{\rho}_b(t) | x + y \rangle\,,
\end{equation}
where $\hat{\rho}_b(t)$ is the reduced bosonic density matrix. For the ideal initial state $|\psi_0\rangle = |0\rangle \otimes |\downarrow\rangle$, the Wigner function
\begin{equation}
\label{W0}
W(x,p,t=0) = \frac{R}{\pi} e^{-R(x^2+p^2)}\,,
\end{equation}

 is Gaussian and positive everywhere. However, negative values are expected to arise during the dynamics~\cite{braunstein2005quantum}, as a consequence of the nonlinear interaction with the qubit. Negative values of Wigner function are physically meaningful features that can be directly measured using unbalanced homodyne optical detection~\cite{smithey1993measurement}, as well as through dispersive coupling to a qubit~\cite{bertet2002direct,wang2009decoherence,wang2016schrodinger}. The Wigner negativity has enabled advanced protocols in quantum computing~\cite{Albarelli2018,Mari2012} and quantum metrology~\cite{liu2022experimental,Albarelli2018}. For simple states, the full Wigner function can alternatively be reconstructed from balanced homodyne measurements~\cite{RevModPhys.81.299} or Fourier transformed from directly measurable characteristic functions~\cite{PhysRevLett.125.043602}.
The negativity volume of the Wigner function $\delta(t)=\frac{1}{2}\int dxdp(|W(x,p,t)|-W(x,p,t))$
is widely considered an overall measure of quantum non-Gaussian interference in phase space \cite{Kenfack_2004}. Another convenient diagnostic is provided by the deepest negative value of the Wigner function at a given time, $W_{min}(t)=|\min[W(x,p,t)]|$~\cite{park2022slowing}. This quantity can be experimentally accessed with significantly fewer resources than a full phase space reconstruction~\cite{wakui2007photon}. Both the negativity volume and the deepest negative value are reliable witnesses of ground-state criticality in spin–boson systems \cite{catrabi,LopezPena2021}, since the mere presence of negativity is an indicator of the transition.

However, as we show below, existing measures of Wigner negativity  do not demonstrate kinks for idealized pure states and unitary dynamics. To overcome this limitation, we start by expressing the bosonic survival probability in terms of the Wigner function and decompose it as
\begin{eqnarray}
\label{Woverlap}
|{\mathcal Z}_b(t)|^2
&=& {\rm Tr}[\hat{\rho}_b(0)\hat{\rho}_b(t)]\nonumber \\
&=& \frac{2\pi}{R}\int dx\,dp\, W_0\,W_t \nonumber\\
&=& P_+(t)-P_-(t),
\end{eqnarray}
where $P_\pm(t)$ are the relative phase space contributions
\begin{equation}
P_+(t)\equiv I_{++}(t)+I_{--}(t),
\qquad
P_-(t)\equiv I_{+-}(t)+I_{-+}(t),
\end{equation}
and
\begin{eqnarray}
\label{Icontributions}
I_{\sigma \sigma'}(t) &=& \sigma \sigma' \frac{2\pi}{R} \int dx\,dp\, W_0W_{t}\,
\Theta\!\big( \sigma W_0\big)\,\Theta\!\big( \sigma' W_t\big), 
\end{eqnarray}
with $\sigma,\sigma' =+\text{ or } -$. Here, $W_0=W(x,p,t=0)$ and $W_t=W(x,p,t)$, while $\Theta(x)$ denotes the Heaviside step function. The quantities \eqref{Icontributions} specify overlaps between positive and negative parts of the initial and evolved Wigner functions. The decomposition of the survival probability \eqref{Woverlap} makes explicit that the cancellation of the survival probability may occur because of the competition between oscillating positive and negative phase space contributions.  If the ratio between the relative phase space contributions $P_-(t_c)/P_+(t_c)\to1$ approaches unity at critical time $t_c$,  the survival probability abruptly vanishes $|{\mathcal Z}_b(t_c)|^2\to 0$, generating a divergence in the rate function \eqref{ratefunction}. Substituting \eqref{Woverlap} into
$\Lambda_b(t)=-(1/R)\log|{\mathcal Z}_b(t)|^2$, the bosonic rate function can be decomposed as
$\Lambda_b(t)=\eta(t)+\Gamma(t)$, where
$\eta(t)\equiv-\frac{1}{R}\log P_+(t)$,
and the second term, which we define as the relative Wigner negativity, is
\begin{equation}
\label{Gamma}
\Gamma(t)\equiv
-\frac{1}{R}\log\!\left(1-\frac{P_-(t)}{P_+(t)}\right).
\end{equation}
The relative Wigner negativity fully relies on the presence of Wigner negativity in either the initial or the evolved state. In the absence of Wigner negativity simultaneously in initial and evolved states, $I_{-+}(t)=I_{+-}(t)=P_-(t)=0$, the relative Wigner negativity vanishes $\Gamma(t)\to 0$. For initial Gaussian states that are positive everywhere like \eqref{W0}, $I_{-+}(t)=I_{--}(t)=0$, the relative Wigner negativity relies entirely on $I_{+-}(t)$, i.e., on the quantum non-Gaussian effects developed during the dynamics.

The decomposition $\Lambda_b(t)=\eta(t)+\Gamma(t)$ reveals two distinct possible origins of the kinks in the bosonic rate function, since any kink in $\Lambda_b(t)$ must necessarily be accompanied by a corresponding kink in either $\eta(t)$ or the relative Wigner negativity $\Gamma(t)$. The argument of $\eta(t)$ is strictly positive and can vanish only in the idealized semiclassical limit $R\to\infty$, where the bosonic state becomes infinitely localized in phase space. In contrast, the argument of $\Gamma(t)$ can approach zero already at finite qubit frequency $R$ when $P_-(t)/P_+(t)\to1$. Thus, the relative negativity $\Gamma(t)$ is the quantum non-Gaussianity quantifier that directly inherits the kinks from the bosonic rate function and identifies quantum non-Gaussian dynamical effects as the mechanism in the phase space. Since singular behavior of $\eta(t)$ is restricted to the semiclassical regime $R\to\infty$, we therefore focus on the relative Wigner negativity $\Gamma(t)$ to identify manifestations of nonequilibrium criticality at finite qubit frequencies $R$.

\begin{figure}[t!]
    \centering
    \includegraphics[width=\columnwidth]{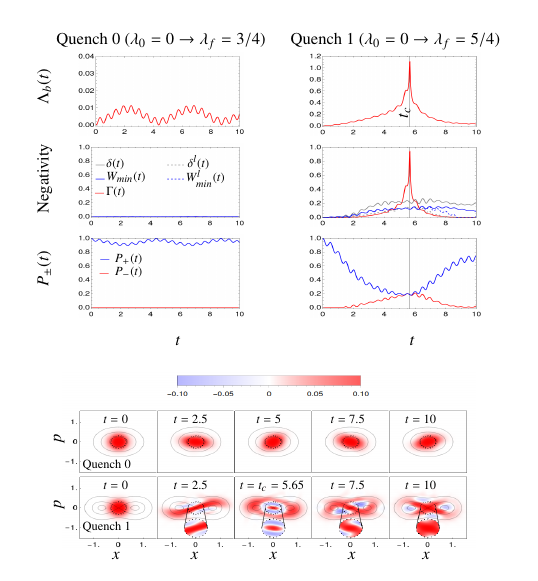}
    \caption{\label{Fig1}Top panel: Bosonic rate function $\Lambda_b(t)$ (first row), Wigner negativity indicators (second row) including negativity volume $\delta(t)$, deepest negative value $W_{min}(t)$ and relative negativity $\Gamma(t)$. The superindex $l$ indicates local quantities within the area occupied by the initial state $x^2+p^2\leq1/R$. Two representative quenches are shown: Quench 0 $(\lambda_0=0\to \lambda_f=3/4)$, with final coupling in the normal phase, and Quench 1 $(\lambda_0=0\to \lambda_f=5/4)$, with final coupling in the superradiant phase. The vertical line indicates the critical time $t_c=5.65$ during Quench 1. Middle panel: Relative phase space contributions $P_\pm(t)$ as a function of the time. Bottom panel: Unitary evolution of the Wigner function. The gray lines denote contours of constant energy $h_{\mathrm{eff}}$. The dotted circle indicates the area occupied by the initial state $x^2+p^2\leq1/R$. For all panels, the qubit frequency is finite $R=10$.}
\end{figure}

To illustrate the kinks from the relative Wigner negativity in the experimentally feasible QRM, we consider two representative quenches: Quench 0 ($\lambda_0 = 0 \to \lambda_f = 3/4$), with final coupling in the normal phase, and Quench 1 ($\lambda_0 = 0 \to \lambda_f = 5/4$), with final coupling in the superradiant phase. For the analysis, we consider the qubit frequency $R=10$ to emphasize that nonequilibrium criticality already manifests at small qubit frequencies, rather than the unphysical semiclassical limit $R\to\infty$. In Fig.~\ref{Fig1}, we show the unitary evolution of the Wigner function together with contours of fixed semiclassical energy $h_{eff}$. We also present the corresponding bosonic rate functions and quantifiers of Wigner negativity. For Quench 0, the energy contours form simple ellipses, and the dynamics remains Gaussian, without Wigner negativity. This regime exhibits smooth oscillations and no kinks in the bosonic rate function. In contrast, for Quench 1, the energy contours develop a double-well structure, leading to strongly non-Gaussian effects during the dynamics. In this case, the bosonic rate function exhibits a clear kink at the critical time $t_c = 5.65$. During Quench 1, neither the Wigner negativity volume $\delta(t)$ nor the minimum value $W_{\min}(t)$ provides a clear signature of criticality, as they do not display any distinctive kink or onset at the critical time. This shows that the mere presence of Wigner negativity is not sufficient to signal nonequilibrium criticality. In contrast, the relative Wigner negativity $\Gamma(t)$ exhibits a clear and sharp kink precisely at the critical time $t_c=5.65$. This correspondence allows us to attribute the nonequilibrium criticality to quantum non-Gaussian effects developed during the dynamics, which enable the strong suppression of the survival probability. This suppression is possible because substantial Wigner negativity develops smoothly within the phase space region occupied by the initial state, such that at the critical time $P_-(t_c)/P_+(t_c)\to 1$. After the critical time, the negativity leaves this region and the survival probability is restored, leading to a smooth decline in both the relative Wigner negativity and the rate function. Unlike other negativity quantifiers, the relative Wigner negativity captures this sharp rise and decline because the negativity is weighted by the initial Wigner function and normalized by the corresponding positive contribution. The logarithmic dependence further amplifies variations in this ratio as it approaches unity, giving rise to the characteristic kink.

The nonanalytic behavior of the bosonic rate function, confirmed through the structure of the Fisher zeros under unitary dynamics, is presented in Appendix~\ref{Fisher}, together with its scaling with the qubit frequency $R$.  In particular, we show that the zeros of the survival probability i.e. the kinks in the bosonic rate function manifest periodically with increasing qubit frequency $\Delta R\approx13$ for Quench 1. Also in Appendix~\ref{Fisher}, we analyze the robustness of the kinks against variations in the quench parameters, specifically the final coupling strength $\lambda_f$.

\begin{figure*}[t!]
    \centering
    \includegraphics[width=\textwidth]{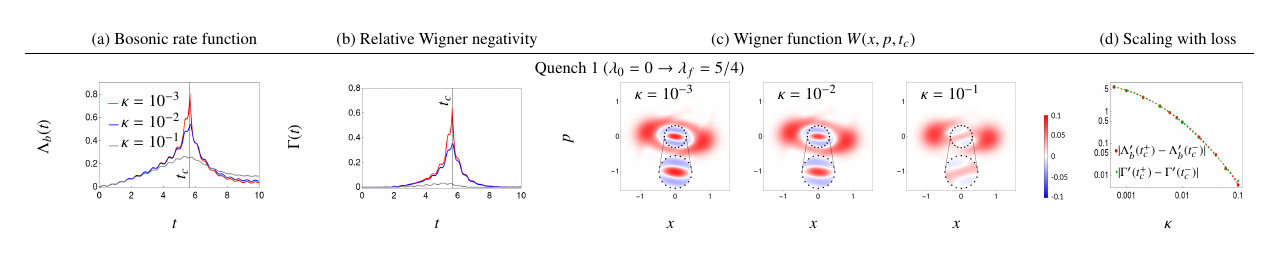}
    \caption{\label{Fig2} (a) Bosonic rate function $\Lambda_b(t)$ as a function of the time $t$. (b) Relative Wigner negativity $\Gamma(t)$. The vertical 
line marks the critical time. (c) Bosonic Wigner function $W(x,p,t_c)$ at the critical time $t_c=5.65$. The dotted circle indicates the area $x^2+p^2\leq1/R$ occupied by the initial state. (d) Scaling with bosonic loss of the discontinuity in the time derivatives of the bosonic rate function $|\Lambda_b^\prime(t_c^+)-\Lambda_b^\prime(t_c^-)|$ and relative Wigner negativity $|\Gamma^\prime(t_c^+)-\Gamma^\prime(t_c^-)|$. The dashed lines are fits of the form $f(\kappa)=e^{-a-b\log \kappa -c (\log \kappa)^2}$.
For all figures the initial state is a thermal state $\hat{\rho}_{th}(\beta=100,\lambda_0=0)$ and  the qubit frequency is finite $R=10$.}
\end{figure*}

\begin{figure}[t!]
    \centering
    \includegraphics[width=\columnwidth]{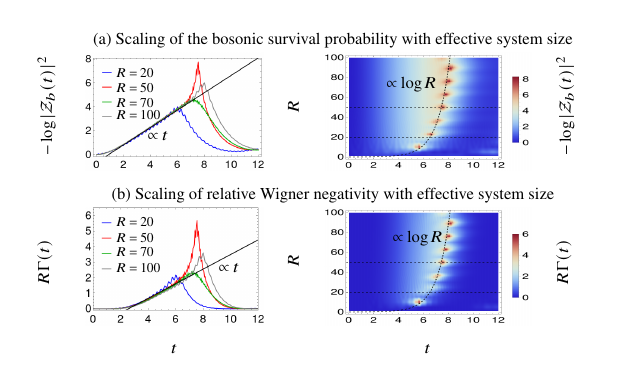}
    \caption{\label{Fig3} (a) Logarithm of the bosonic survival probability $-\log|{\mathcal Z}_b(t)|^2=R\Lambda_b(t)$ without rescaling, with initial thermal occupation $\hat{\rho}_{th}(\beta=100,\lambda_0=0)$ and under loss $\kappa=10^{-3}$, for Quench 1 $(\lambda_0=0\to\lambda_f=5/4)$. On the left column, we show the initial linear growth with time, as illustrated by the black line. On the right column, the logarithm of the survival probability is shown as a function of time $t$ and qubit frequency $R$. The intense red regions corresponds to the zeros that produce kinks in the bosonic rate function $\Lambda_b(t)$. They reappear periodically after approximately $\Delta R\approx13$. The black dashed line indicates the $\log R$ shift of the zeros. The horizontal lines mark $R=50$ and $R=20$. (b) Same for the relative Wigner negativity $R\Gamma(t)$ without rescaling.}
\end{figure}

To assess robustness under experimental conditions and fulfill point (iii), we consider the extension to mixed-state dynamics. In particular, the initial state can be generalized to a Gibbs thermal state of the QRM, which naturally arises as the steady state in thermal equilibrium \cite{PhysRevResearch.6.013001}. Its density matrix is given by
\begin{equation}
\label{thermalstate}
\hat{\rho}_{th}(\beta,\lambda_0) = \frac{e^{-\beta \hat{H}(\lambda_0)}}{\sum_k e^{-\beta E_k(\lambda_0)}},
\end{equation}
where $\beta=1/T$ is the inverse of the temperature. Following the quench, the initial thermal state evolves under open-system dynamics, with loss included as a common source of environmental decoherence in bosonic systems \cite{gu2017microwave,Mirrahimi_2014}. The time evolution of the density matrix is then governed by the Lindblad master equation \cite{breuer2002theory}
\begin{equation}
\label{Lindblad}
\frac{d\hat{\rho}(t)}{dt}
=
-i[\hat{H}(\lambda_f),\hat{\rho}(t)]
+ \kappa\, D[\hat{a}]\,\hat{\rho}(t),
\end{equation}
where $\kappa$ denotes the bosonic loss rate and the Lindblad dissipator is $
D[\hat{O}]\,\hat{\rho}
=
\hat{O}\hat{\rho}\hat{O}^\dagger
-\frac{1}{2}
\left\{
\hat{O}^\dagger\hat{O},\,\hat{\rho}
\right\}.$
For mixed-state dynamics under weak bosonic loss $\kappa \ll 1$ and low thermal occupation $T \gtrsim 0$, the bosonic survival probability remains a good approximation to the full spin--boson survival probability, $|{\mathcal Z}(t)|^2 \approx |{\mathcal Z}_b(t)|^2$. Indeed, the overlap between the initial and the evolved density matrices is one of the proposed extensions of the survival probability for non-unitary dynamics under dissipation \cite{Link2020} (a comparison with other possible extensions is presented in Appendix \ref{surprob}). Consequently, the kinks in the bosonic rate function $\Lambda_b(t)$ are still sufficient diagnostic of nonequilibrium criticality in the bosonic system.
In Fig.~\ref{Fig2}, we show that the kinks in the bosonic rate function and in relative Wigner negativity persist at finite temperature and under weak bosonic loss. Only when Wigner negativity is substantially suppressed by strong dissipation do the kinks become smooth and the criticality disappears. Remarkably, the sharpness of the kinks in the bosonic rate function and the relative Wigner negativity exhibit essentially the same scaling with bosonic loss, as quantified by the discontinuities in their time derivatives. This close correspondence provides further evidence that the kinks are intrinsically tied to the quantum non-Gaussian effects developed during the dynamics. To further substantiate the phase space origin of the kinks as a consequence of the  quantum non-Gaussian effects, we analyze approximate descriptions of the dynamics, including the quantization of the semiclassical Hamiltonian \eqref{Hsemiclassical} and the Truncated Wigner approximation (TWA), in Appendix \ref{correspondence}. These complementary approaches collectively demonstrate that these kinks are genuine quantum non-Gaussian phenomena that cannot be captured in the absence of Wigner negativity. In addition, we investigate the depth of the relative Wigner negativity in Appendix \ref{Depht}, showing that the bosonic rate function becomes smooth when the relative Wigner negativity is reduced by Gaussian smoothing noise. To verify feasible scaling, in Fig.~\ref{Fig3} we show that under loss, the kinks in both $\Lambda_{b}$ and $\Gamma(t)$ still appear
periodically with increasing qubit frequency.

Beyond the quenches where initial states are Gaussian, we also examined initial ground states in the superradiant phase $\lambda_0>\lambda_c$, where the qubit and bosonic mode are entangled \cite{catrabi} and the initial Wigner function already contains negative regions (see Appendix~\ref{oppositequench}). For these quenches, the correspondence between kinks in the bosonic rate function and the relative Wigner negativity persists, indicating that they are still tied to the quantum non-Gaussian effects. Furthermore, the kinks in the bosonic rate function are more robust against bosonic loss because the negativity already present in the initial state is not affected by the subsequent loss, which only suppresses the Wigner negativity generated during the dynamics. The kinks in the full spin--boson rate function also persist, although they are slightly shifted.

All the evidence presented here indicates that quantum non-Gaussian dynamical effects allow abrupt suppression of the bosonic survival probability through the competition between the relative phase space contributions. This abrupt suppression of the bosonic survival probability is the phase space mechanism behind the kinks in the bosonic rate function at finite qubit frequencies. The relative Wigner negativity is the quantum non-Gaussianity quantifier that develops kinks during the nonequilibrium criticality  with adequate scalings in time and system size. In this sense, the relative Wigner negativity brings phase space understanding of the kinks in the bosonic rate function generated by quantum non-Gaussian effects, identifying them as genuinely quantum phenomena beyond the reach of semiclassical descriptions. The characteristic kinks in the relative Wigner negativity remain robust under weak bosonic loss and finite thermal occupation. The kinks are correspondingly smoothed as dissipation becomes sufficiently strong to suppress the Wigner negativity responsible for the phase space cancellation.

The kinks in the Wigner negativity as realistic manifestations of DQPT in the idealized setting are expected to manifest in other nonlinear bosonic systems, such as Kerr oscillators \cite{Prado2023,kmvc-88rh,PhysRevA.102.063531,doi:10.1073/pnas.2311241121}, where reconstruction of the Wigner function is also experimentally accessible on different platforms \cite{marti2024kerr,he2023fast}. Furthermore, the decomposition of the survival probability and rate function can be extended to discrete-variable systems by employing the SU(2) Wigner function \cite{davis2021wigner,klimov20022wignersu2,akhtar2021sub}. Such an extension represents a promising direction for future research and will be explored elsewhere.

{\it Acknowledgments.}
D.J.N. and R. F. acknowledge the support from grant CZ.02.01.01/00/22\_008/0004649
(QUEENTEC) provided by the EU and MEYS of the Czech Republic. D.J.N. also acknowledges QuantERA project Q3ML (9C26002) of the MEYS of the Czech Republic. R.F. was also supported by the QuantERA project CLUSSTAR (8C2024003) of the MEYS of the Czech Republic. Projects
CLUSSTAR and Q3ML have received funding from the EU Horizon Programme under Grant Agreement No. 731473 and 101017733
(QuantERA). 
P. C. , P. S. and J. N.  acknowledge grant 25-16056S of the Czech Science Foundation.

{\it Data Availability.} All codes and scripts needed to reproduce the results in this paper are available online \url{https://github.com/djuliannader/RabiModel}.
 The time evolution of the quantum state was obtained by numerically diagonalizing the Hamiltonian~\eqref{HQRM} in the tensor-product basis of the qubit states and the oscillator Fock states. For all results, convergence with respect to the Fock-basis truncation was confirmed by ensuring that the cutoff $N$ satisfies
 $\int_0^{10}||{\mathcal Z}_b^{(N+10)}(t)|^2-|{\mathcal Z}_b^{(N)}(t)|^2|dt<10^{-6}\,.$
 The survival amplitude ${\mathcal Z}(t)$ and its analytic continuation to complex time were obtained directly from the eigenspectrum and overlaps with the initial state. The Wigner function~\eqref{Wignerqpt} was computed using the \textit{QuantumOptics.jl} framework~\cite{QuantumOptics}. The Lindblad master equation \eqref{Lindblad} describing the dynamics under loss was solved numerically using the Julia package for differential equations \textit{DifferentialEquations.jl}
\cite{rackauckas2017differentialequations}. The unitary dynamics under the effective Hamiltonian \eqref{Hqsemiclassical} was obtained using finite-difference discretization \cite{NaderBEC2025}. The Langevin equations \eqref{Langevineq} necessary for the TWA were solved numerically in Mathematica using time Euler–Maruyama discretization \cite{Higham2001}.

\bibliography{Bibliography.bib}

@article{akhtar2021sub,
  title={Sub-{P}lanck structures: Analogies between the {H}eisenberg-{W}eyl and {SU}(2) groups},
  author={Akhtar, Naeem and Sanders, Barry C and Navarrete-Benlloch, Carlos},
  journal={Phys. Rev. A.},
  volume={103},
  number={5},
  pages={053711},
  year={2021},
  publisher={APS},
  doi = {10.1103/PhysRevA.103.053711},
  url = {https://link.aps.org/doi/10.1103/PhysRevA.103.053711}
}

@article{adesso2014continuous,
  title={Continuous variable quantum information: Gaussian states and beyond},
  author={Adesso, Gerardo and Ragy, Sammy and Lee, Antony R},
  journal={Open Systems \& Information Dynamics},
  volume={21},
  number={01n02},
  pages={1440001},
  year={2014},
  publisher={World Scientific},
  URL = {https://doi.org/10.1142/S1230161214400010},
}

@article{Albarelli2018,
  title = {Resource theory of quantum non-{G}aussianity and {W}igner negativity},
  author = {Albarelli, Francesco and Genoni, Marco G. and Paris, Matteo G. A. and Ferraro, Alessandro},
  journal = {Phys. Rev. A},
  volume = {98},
  issue = {5},
  pages = {052350},
  numpages = {17},
  year = {2018},
  month = {Nov},
  publisher = {American Physical Society},
  doi = {10.1103/PhysRevA.98.052350},
  url = {https://link.aps.org/doi/10.1103/PhysRevA.98.052350}
}

@article{Kenfack_2004,
doi = {10.1088/1464-4266/6/10/003},
url = {https://dx.doi.org/10.1088/1464-4266/6/10/003},
year = {2004},
month = {aug},
publisher = {},
volume = {6},
number = {10},
pages = {396},
author = {Anatole Kenfack and Karol Życzkowski},
title = {Negativity of the {W}igner function as an indicator of non-classicality},
journal = { J. Opt. B: Quantum Semiclass. Opt.}
}

@book{breuer2002theory,
  title={The theory of open quantum systems},
  author={Breuer, Heinz-Peter and Petruccione, Francesco},
  year={2002},
  publisher={Oxford University Press, USA},
  url = {https://doi.org/10.1093/acprof:oso/9780199213900.001.0001}
}

@article{bertet2002direct,
  title={Direct measurement of the {W}igner function of a one-photon {F}ock state in a cavity},
  author={Bertet, Patrice and Auffeves, Alexia and Maioli, Paolo and Osnaghi, Stefano and Meunier, Tristan and Brune, Michel and Raimond, Jean-Michel and Haroche, Serge},
  journal={Phys. Rev. Lett. },
  volume={89},
  number={20},
  pages={200402},
  year={2002},
  publisher={APS},
  doi = {10.1103/PhysRevLett.89.200402},
  url = {https://link.aps.org/doi/10.1103/PhysRevLett.89.200402}
}

@article{braunstein2005quantum,
  title={Quantum information with continuous variables},
  author={Braunstein, Samuel L and Van Loock, Peter},
  journal={Rev. Mod. Phys.},
  volume={77},
  number={2},
  pages={513--577},
  year={2005},
  publisher={APS},
  doi = {10.1103/RevModPhys.77.513},
  url = {https://link.aps.org/doi/10.1103/RevModPhys.77.513}
}

@article{braumuller2017analog,
  title={Analog quantum simulation of the {R}abi model in the ultra-strong coupling regime},
  author={Braum{\"u}ller, Jochen and Marthaler, Michael and Schneider, Andre and Stehli, Alexander and Rotzinger, Hannes and Weides, Martin and Ustinov, Alexey V},
  journal={Nat. Commun.},
  volume={8},
  number={1},
  pages={779},
  year={2017},
  publisher={Nature Publishing Group UK London},
  url={https://doi.org/10.1038/s41467-017-00894-w}
}

@article{bhattacharya2017mixed,
  title={Mixed state dynamical quantum phase transitions},
  author={Bhattacharya, Utso and Bandyopadhyay, Souvik and Dutta, Amit},
  journal={Phys. Rev. B},
  volume={96},
  number={18},
  pages={180303},
  year={2017},
  publisher={APS},
  doi = {10.1103/PhysRevB.96.180303},
  url = {https://link.aps.org/doi/10.1103/PhysRevB.96.180303}
}

@article{Braak2011,
  title = {Integrability of the {R}abi Model},
  author = {Braak, D.},
  journal = {Phys. Rev. Lett.},
  volume = {107},
  issue = {10},
  pages = {100401},
  numpages = {4},
  year = {2011},
  month = {Aug},
  publisher = {American Physical Society},
  doi = {10.1103/PhysRevLett.107.100401},
  url = {https://link.aps.org/doi/10.1103/PhysRevLett.107.100401}
}

@article{borrmann2000classification,
  title={Classification of phase transitions in small systems},
  author={Borrmann, Peter and M{\"u}lken, Oliver and Harting, Jens},
  journal={Phys. Rev. Lett. },
  volume={84},
  number={16},
  pages={3511},
  year={2000},
  publisher={APS},
  doi = {10.1103/PhysRevLett.84.3511},
  url = {https://link.aps.org/doi/10.1103/PhysRevLett.84.3511}
}

@article{bian2026temporal,
  title={Temporal Dynamical Quantum Phase Transition in Dicke Model with Trapped Ions},
  author={Bian, Ji and Wu, Wei and Xie, Zihan and Zhang, Mengxiang and Li, Yi and Li, Yue and Yao, Rixin and Zhou, Yuqi and Cheng, Xu and Pu, Han and others},
  journal={arXiv preprint arXiv:2606.28664},
  year={2026}
}

@article{CEJNAR2009210,
title = {Quantum phase transitions in the interacting boson model},
journal = {Progress in Particle and Nuclear Physics},
volume = {62},
number = {1},
pages = {210-256},
year = {2009},
issn = {0146-6410},
doi = {https://doi.org/10.1016/j.ppnp.2008.08.001},
url = {https://www.sciencedirect.com/science/article/pii/S0146641008000719},
author = {Pavel Cejnar and Jan Jolie}
}

@book{carr2010understanding,
  title={Understanding quantum phase transitions},
  author={Carr, Lincoln},
  year={2010},
  publisher={CRC press},
  url={https://doi.org/10.1201/b10273}
}

@article{corney2015non,
  title = {Non-{G}aussian pure states and positive {W}igner functions},
  author = {Corney, J. F. and Olsen, M. K.},
  journal = {Phys. Rev. A},
  volume = {91},
  issue = {2},
  pages = {023824},
  numpages = {6},
  year = {2015},
  month = {Feb},
  publisher = {American Physical Society},
  doi = {10.1103/PhysRevA.91.023824},
  url = {https://link.aps.org/doi/10.1103/PhysRevA.91.023824}
}

@article{Chen2020,
  title = {Experimentally detecting dynamical quantum phase transitions in a slowly quenched {I}sing-chain model},
  author = {Chen, Zhe and Cui, Jin-Ming and Ai, Ming-Zhong and He, Ran and Huang, Yun-Feng and Han, Yong-Jian and Li, Chuan-Feng and Guo, Guang-Can},
  journal = {Phys. Rev. A},
  volume = {102},
  issue = {4},
  pages = {042222},
  numpages = {9},
  year = {2020},
  month = {Oct},
  publisher = {American Physical Society},
  doi = {10.1103/PhysRevA.102.042222},
  url = {https://link.aps.org/doi/10.1103/PhysRevA.102.042222}
}

@article{catrabi,
  title = {Using nonclassical states to explore the superradiant phase transition in the quantum {R}abi model},
  author = {Liu, Junpeng and Zhao, Miaomiao and Yang, Yun-Tong and Luo, Hong-Gang},
  journal = {Phys. Rev. A},
  volume = {109},
  issue = {2},
  pages = {023721},
  numpages = {12},
  year = {2024},
  month = {Feb},
  publisher = {American Physical Society},
  doi = {10.1103/PhysRevA.109.023721},
  url = {https://link.aps.org/doi/10.1103/PhysRevA.109.023721}
}

@article{caprio2008excited,
title = {Excited state quantum phase transitions in many-body systems},
journal = {Ann. Phys.},
volume = {323},
number = {5},
pages = {1106-1135},
year = {2008},
issn = {0003-4916},
doi = {https://doi.org/10.1016/j.aop.2007.06.011},
url = {https://www.sciencedirect.com/science/article/pii/S0003491607001042},
author = {M.A. Caprio and P. Cejnar and F. Iachello}
}

@article{cejnar2006monodromy,
doi = {10.1088/0305-4470/39/31/L01},
url = {https://doi.org/10.1088/0305-4470/39/31/L01},
year = {2006},
month = {jul},
publisher = {},
volume = {39},
number = {31},
pages = {L515},
author = {Cejnar, Pavel and Macek, Michal and Heinze, Stefan and Jolie, Jan and Dobeš, Jan},
title = {Monodromy and excited-state quantum phase transitions in integrable systems: collective vibrations of nuclei},
journal = {J. Phys. A: Math. Gen.}
}

@article{cejnar2021excited,
doi = {10.1088/1751-8121/abdfe8},
url = {https://doi.org/10.1088/1751-8121/abdfe8},
year = {2021},
month = {mar},
publisher = {IOP Publishing},
volume = {54},
number = {13},
pages = {133001},
author = {Cejnar, Pavel and Stránský, Pavel and Macek, Michal and Kloc, Michal},
title = {Excited-state quantum phase transitions},
journal = {J. Phys. A: Math. Theor.}
}

@article{complextime,
  title = {Mechanism of dynamical phase transitions: The complex-time survival amplitude},
  author = {Corps, \'Angel L. and Str\'ansk\'y, Pavel and Cejnar, Pavel},
  journal = {Phys. Rev. B},
  volume = {107},
  issue = {9},
  pages = {094307},
  numpages = {15},
  year = {2023},
  month = {Mar},
  publisher = {American Physical Society},
  doi = {10.1103/PhysRevB.107.094307},
  url = {https://link.aps.org/doi/10.1103/PhysRevB.107.094307}
}

@article{liu2024quantum,
  title={Quantum phase transition of the {J}aynes-{C}ummings model},
  author={Liu, Cheng and Huang, Jin-Feng},
  journal={Sci. China-Phys. Mech. Astron.},
  volume={67},
  number={1},
  pages={210311},
  year={2024},
  publisher={Springer},
  url = {https://doi.org/10.1007/s11433-023-2243-7}
}

@article{cai2021observation,
  title={Observation of a quantum phase transition in the quantum {R}abi model with a single trapped ion},
  author={Cai, M-L and Liu, Z-D and Zhao, W-D and Wu, Y-K and Mei, Q-X and Jiang, Y and He, L and Zhang, X and Zhou, Z-C and Duan, L-M},
  journal={Nat. Commun.},
  volume={12},
  number={1},
  pages={1126},
  year={2021},
  publisher={Nature Publishing Group UK London},
  doi = {10.1038/s41467-021-21425-8}
}

@article{crespi2012photonic,
  title={Photonic realization of the quantum {R}abi model},
  author={Crespi, Andrea and Longhi, Stefano and Osellame, Roberto},
  journal={Phys. Rev. Lett.},
  volume={108},
  number={16},
  pages={163601},
  year={2012},
  publisher={APS},
  doi = {10.1103/PhysRevLett.108.163601},
  url = {https://link.aps.org/doi/10.1103/PhysRevLett.108.163601}
}

@article{davis2021wigner,
  title={Wigner negativity in spin-j systems},
  author={Davis, Jack and Kumari, Meenu and Mann, Robert B and Ghose, Shohini},
  journal={Phys. Rev. Res.},
  volume={3},
  number={3},
  pages={033134},
  year={2021},
  publisher={APS},
  doi = {10.1103/PhysRevResearch.3.033134},
  url = {https://link.aps.org/doi/10.1103/PhysRevResearch.3.033134}
}

@article{divakaran2016tuning,
  title={Tuning the presence of dynamical phase transitions in a generalized XY spin chain},
  author={Divakaran, Uma and Sharma, Shraddha and Dutta, Amit},
  journal={Phys. Rev. E},
  volume={93},
  number={5},
  pages={052133},
  year={2016},
  publisher={APS},
  doi = {10.1103/PhysRevE.93.052133},
  url = {https://link.aps.org/doi/10.1103/PhysRevE.93.052133}
}

@article{derrida1991zeroes,
  title={The zeroes of the partition function of the random energy model},
  author={Derrida, Bernard},
  journal={Physica A},
  volume={177},
  number={1-3},
  pages={31--37},
  year={1991},
  publisher={Elsevier},
  doi = {https://doi.org/10.1016/0378-4371(91)90130-5}
}

@article{TWA,
doi = {10.1088/0305-4470/13/7/018},
url = {https://dx.doi.org/10.1088/0305-4470/13/7/018},
year = {1980},
month = {jul},
publisher = {},
volume = {13},
number = {7},
pages = {2353},
author = {P D Drummond and  C W Gardiner},
title = {Generalised P-representations in quantum optics},
journal = {Journal of Physics A: Mathematical and General}
}

@article{forn2010observation,
  title={Observation of the {B}loch-{S}iegert Shift in a Qubit-Oscillator System in the Ultrastrong Coupling Regime},
  author={Forn-D{\'\i}az, Pol and Lisenfeld, J{\"u}rgen and Marcos, David and Garcia-Ripoll, Juan Jos{\'e} and Solano, Enrique and Harmans, CJPM and Mooij, JE},
  journal={Phys. Rev. Lett.},
  volume={105},
  number={23},
  pages={237001},
  year={2010},
  publisher={APS},
  doi = {10.1103/PhysRevLett.105.237001},
  url = {https://link.aps.org/doi/10.1103/PhysRevLett.105.237001}
}

@article{filip2002overlap,
  title={Overlap and entanglement-witness measurements},
  author={Filip, Radim},
  journal={Phy. Rev. A},
  volume={65},
  number={6},
  pages={062320},
  year={2002},
  publisher={APS},
  doi = {10.1103/PhysRevA.65.062320},
  url = {https://link.aps.org/doi/10.1103/PhysRevA.65.062320}
}

@article{PhysRevLett.125.043602,
  title = {Direct Characteristic-Function Tomography of Quantum States of the Trapped-Ion Motional Oscillator},
  author = {Fl\"uhmann, C. and Home, J. P.},
  journal = {Phys. Rev. Lett.},
  volume = {125},
  issue = {4},
  pages = {043602},
  numpages = {6},
  year = {2020},
  month = {Jul},
  publisher = {American Physical Society},
  doi = {10.1103/PhysRevLett.125.043602},
  url = {https://link.aps.org/doi/10.1103/PhysRevLett.125.043602}
}

@article{PhysRevB.96.104436,
  title = {Anomalous dynamical phase in quantum spin chains with long-range interactions},
  author = {Homrighausen, Ingo and Abeling, Nils O. and Zauner-Stauber, Valentin and Halimeh, Jad C.},
  journal = {Phys. Rev. B},
  volume = {96},
  issue = {10},
  pages = {104436},
  numpages = {7},
  year = {2017},
  month = {Sep},
  publisher = {American Physical Society},
  doi = {10.1103/PhysRevB.96.104436},
  url = {https://link.aps.org/doi/10.1103/PhysRevB.96.104436}
}

@article{hillery1984distribution,
  title={Distribution functions in physics: Fundamentals},
  author={Hillery, MOSM and O'Connell, Robert F and Scully, Marlan O and Wigner, Eugene P},
  journal={Phys. Rep.},
  volume={106},
  number={3},
  pages={121--167},
  year={1984},
  publisher={Elsevier},
  doi = {https://doi.org/10.1016/0370-1573(84)90160-1},
}

@article{irish2014oscillator,
  title={Oscillator tunneling dynamics in the {R}abi model},
  author={Irish, EK and Gea-Banacloche, Julio},
  journal={Phys. Rev. B. },
  volume={89},
  number={8},
  pages={085421},
  year={2014},
  publisher={APS},
  doi = {10.1103/PhysRevB.89.085421},
  url = {https://link.aps.org/doi/10.1103/PhysRevB.89.085421}
}

@article{
doi:10.1073/pnas.2311241121,
author = {Jayameenakshi Venkatraman  and Rodrigo G. Cortiñas  and Nicholas E. Frattini  and Xu Xiao  and Michel H. Devoret },
title = {A driven {K}err oscillator with two-fold degeneracies for qubit protection},
journal = {Proceedings of the National Academy of Sciences},
volume = {121},
number = {24},
pages = {e2311241121},
year = {2024},
doi = {10.1073/pnas.2311241121},
URL = {https://www.pnas.org/doi/abs/10.1073/pnas.2311241121},
eprint = {https://www.pnas.org/doi/pdf/10.1073/pnas.2311241121}}

@article{DQPTtrappedions,
  title = {Direct Observation of Dynamical Quantum Phase Transitions in an Interacting Many-Body System},
  author = {Jurcevic, P. and Shen, H. and Hauke, P. and Maier, C. and Brydges, T. and Hempel, C. and Lanyon, B. P. and Heyl, M. and Blatt, R. and Roos, C. F.},
  journal = {Phys. Rev. Lett.},
  volume = {119},
  issue = {8},
  pages = {080501},
  numpages = {5},
  year = {2017},
  month = {Aug},
  publisher = {American Physical Society},
  doi = {10.1103/PhysRevLett.119.080501},
  url = {https://link.aps.org/doi/10.1103/PhysRevLett.119.080501}
}

@article{RevModPhys.81.299,
  title = {Continuous-variable optical quantum-state tomography},
  author = {Lvovsky, A. I. and Raymer, M. G.},
  journal = {Rev. Mod. Phys.},
  volume = {81},
  issue = {1},
  pages = {299--332},
  numpages = {0},
  year = {2009},
  month = {Mar},
  publisher = {American Physical Society},
  doi = {10.1103/RevModPhys.81.299},
  url = {https://link.aps.org/doi/10.1103/RevModPhys.81.299}
}

@article{lang2018dynamical,
  title={Dynamical quantum phase transition for mixed states in open systems},
  author={Lang, Haifeng and Chen, Yixin and Hong, Qiantan and Fan, Heng},
  journal={Phy. Rev. B},
  volume={98},
  number={13},
  pages={134310},
  year={2018},
  publisher={APS},
  doi = {10.1103/PhysRevB.98.134310},
  url = {https://link.aps.org/doi/10.1103/PhysRevB.98.134310}
}

@article{Lee1991,
  title = {Measure of the nonclassicality of nonclassical states},
  author = {Lee, Ching Tsung},
  journal = {Phys. Rev. A},
  volume = {44},
  issue = {5},
  pages = {R2775--R2778},
  numpages = {0},
  year = {1991},
  month = {Sep},
  publisher = {American Physical Society},
  doi = {10.1103/PhysRevA.44.R2775},
  url = {https://link.aps.org/doi/10.1103/PhysRevA.44.R2775}
}

@article{Liu2024,
  title = {Exact Fisher zeros and thermofield dynamics across a quantum critical point},
  author = {Liu, Yang and Lv, Songtai and Meng, Yuchen and Tan, Zefan and Zhao, Erhai and Zou, Haiyuan},
  journal = {Phys. Rev. Res.},
  volume = {6},
  issue = {4},
  pages = {043139},
  numpages = {6},
  year = {2024},
  month = {Nov},
  publisher = {American Physical Society},
  doi = {10.1103/PhysRevResearch.6.043139},
  url = {https://link.aps.org/doi/10.1103/PhysRevResearch.6.043139}
}

@article{lerma2018analytical,
  title={Analytical description of the survival probability of coherent states in regular regimes},
  author={Lerma-Hern{\'a}ndez, Sergio and Ch{\'a}vez-Carlos, Jorge and Bastarrachea-Magnani, Miguel A and Santos, Lea F and Hirsch, Jorge G},
  journal={J. Phys. A: Math. Theor. },
  volume={51},
  number={47},
  pages={475302},
  year={2018},
  publisher={IOP Publishing},
  doi = {10.1088/1751-8121/aae2c3}
}

@Article{photonics10111184,
AUTHOR = {Li, Ming and Wang, Yinuo and Song, Zhaoyang and Zhao, Yiming and Zhao, Xiaolong and Ma, Hongyang},
TITLE = {Dynamics Reflects Quantum Phase Transition of Rabi Model},
JOURNAL = {Photonics},
VOLUME = {10},
YEAR = {2023},
NUMBER = {11},
ARTICLE-NUMBER = {1184},
URL = {https://www.mdpi.com/2304-6732/10/11/1184},
ISSN = {2304-6732},
DOI = {10.3390/photonics10111184}
}

@article{fisher1967theory,
  title={The theory of equilibrium critical phenomena},
  author={Fisher, Michael E},
  journal={Rep. Prog. Phys.},
  volume={30},
  number={2},
  pages={615},
  year={1967},
  publisher={IOP Publishing},
  doi = {10.1088/0034-4885/30/2/306}
}

@article{gu2017microwave,
  title={Microwave photonics with superconducting quantum circuits},
  author={Gu, Xiu and Kockum, Anton Frisk and Miranowicz, Adam and Liu, Yu-xi and Nori, Franco},
  journal={Phys. Rep.},
  volume={718-719},
  pages={1--102},
  year={2017},
  doi={10.1016/j.physrep.2017.10.002}
}

@article{grossmann1969phase,
  title={Phase transitions and the distribution of temperature zeros of the partition function: I. General relations},
  author={Grossmann, S and Rosenhauer, W},
  journal={Z Physik},
  volume={218},
  number={5},
  pages={437--448},
  year={1969},
  publisher={Springer},
  url = {https://doi.org/10.1007/BF01392423}
}

@article{gunter2009sub,
  title={Sub-cycle switch-on of ultrastrong light--matter interaction},
  author={G{\"u}nter, G and Anappara, Aji A and Hees, J and Sell, Alexander and Biasiol, Giorgio and Sorba, Lucia and De Liberato, S and Ciuti, Cristiano and Tredicucci, Alessandro and Leitenstorfer, Alfred and others},
  journal={Nature},
  volume={458},
  number={7235},
  pages={178--181},
  year={2009},
  publisher={Nature Publishing Group UK London},
  url= {https://doi.org/10.1038/nature07838}
}

@book{gardiner2004quantum,
  title={Quantum Noise: A Handbook of Markovian and Non-Markovian Quantum Stochastic Methods with Applications to Quantum Optics},
  author={Alberto Galindo and Pedro Pascual},
  year={2004},
  url={https://api.semanticscholar.org/CorpusID:122444575}
}

@article{Guo2019,
  title = {Observation of a Dynamical Quantum Phase Transition by a Superconducting Qubit Simulation},
  author = {Guo, Xue-Yi and Yang, Chao and Zeng, Yu and Peng, Yi and Li, He-Kang and Deng, Hui and Jin, Yi-Rong and Chen, Shu and Zheng, Dongning and Fan, Heng},
  journal = {Phys. Rev. Appl.},
  volume = {11},
  issue = {4},
  pages = {044080},
  numpages = {12},
  year = {2019},
  month = {Apr},
  publisher = {American Physical Society},
  doi = {10.1103/PhysRevApplied.11.044080},
  url = {https://link.aps.org/doi/10.1103/PhysRevApplied.11.044080}
}

@article{Gan2020,
  title = {Hybrid Quantum Computing with Conditional Beam Splitter Gate in Trapped Ion System},
  author = {Gan, H. C. J. and Maslennikov, Gleb and Tseng, Ko-Wei and Nguyen, Chihuan and Matsukevich, Dzmitry},
  journal = {Phys. Rev. Lett.},
  volume = {124},
  issue = {17},
  pages = {170502},
  numpages = {6},
  year = {2020},
  month = {Apr},
  publisher = {American Physical Society},
  doi = {10.1103/PhysRevLett.124.170502},
  url = {https://link.aps.org/doi/10.1103/PhysRevLett.124.170502}
}

@article{Gao2018,
  title = {Programmable Interference between Two Microwave Quantum Memories},
  author = {Gao, Yvonne Y. and Lester, Brian J. and Zhang, Yaxing and Wang, Chen and Rosenblum, Serge and Frunzio, Luigi and Jiang, Liang and Girvin, S. M. and Schoelkopf, Robert J.},
  journal = {Phys. Rev. X},
  volume = {8},
  issue = {2},
  pages = {021073},
  numpages = {7},
  year = {2018},
  month = {Jun},
  publisher = {American Physical Society},
  doi = {10.1103/PhysRevX.8.021073},
  url = {https://link.aps.org/doi/10.1103/PhysRevX.8.021073}
}

@article{Higham2001,
  author  = {Higham, Desmond J.},
  title   = {An Algorithmic Introduction to Numerical
             Simulation of Stochastic Differential Equations},
  journal = {SIAM Review},
  volume  = {43},
  number  = {3},
  pages   = {525--546},
  year    = {2001},
  doi     = {10.1137/S0036144500378302}
}

@article{Heyl2017DynamicalQP,
  title={Dynamical quantum phase transitions: a review},
  author={Markus Heyl},
  journal={Rep. Prog. Phys.},
  year={2017},
  volume={81},
  url={https://api.semanticscholar.org/CorpusID:46814318}
}

@article{Heyl2013,
  title = {Dynamical Quantum Phase Transitions in the Transverse-Field Ising Model},
  author = {Heyl, M. and Polkovnikov, A. and Kehrein, S.},
  journal = {Phys. Rev. Lett.},
  volume = {110},
  issue = {13},
  pages = {135704},
  numpages = {5},
  year = {2013},
  month = {Mar},
  publisher = {American Physical Society},
  doi = {10.1103/PhysRevLett.110.135704},
  url = {https://link.aps.org/doi/10.1103/PhysRevLett.110.135704}
}

@article{he2023fast,
  title={Fast generation of {S}chr{\"o}dinger cat states using a {K}err-tunable superconducting resonator},
  author={He, XL and Lu, Yong and Bao, DQ and Xue, Hang and Jiang, WB and Wang, Zhen and Roudsari, AF and Delsing, Per and Tsai, JS and Lin, ZR},
  journal={Nat Commun},
  volume={14},
  number={1},
  pages={6358},
  year={2023},
  publisher={Nature Publishing Group UK London},
  url = {https://doi.org/10.1038/s41467-023-42057-0}
}

@article{Bojan2018,
  title = {Dynamical Quantum Phase Transitions in Spin Chains with Long-Range Interactions: Merging Different Concepts of Nonequilibrium Criticality},
  author = {{\ifmmode \check{Z}\else \v{Z}\fi{}}unkovi{\ifmmode \check{c}\else \v{c}\fi{}}, Bojan and Heyl, Markus and Knap, Michael and Silva, Alessandro},
  journal = {Phys. Rev. Lett.},
  volume = {120},
  issue = {13},
  pages = {130601},
  numpages = {6},
  year = {2018},
  month = {Mar},
  publisher = {American Physical Society},
  doi = {10.1103/PhysRevLett.120.130601},
  url = {https://link.aps.org/doi/10.1103/PhysRevLett.120.130601}
}

@article{jaynes1963comparison,
  title={Comparison of quantum and semiclassical radiation theories with application to the beam maser},
  author={Jaynes, Edwin T and Cummings, Frederick W},
  journal={Proceedings of the IEEE},
  volume={51},
  number={1},
  pages={89--109},
  year={1963},
  publisher={IEEE},
  url={https://api.semanticscholar.org/CorpusID:119713485}
}

@article{klimov20022wignersu2,
  title={On the SU (2) {W}igner function dynamics},
  author={Klimov, AB and Chumakov, SM},
  journal={Rev. Mex. Fis.},
  volume={48},
  number={4},
  pages={317--324},
  year={2002},
  publisher={Sociedad mexicana de f{\'\i}sica},
  url = {http://www.scielo.org.mx/scielo.php?script=sci_arttext&pid=S0035-001X2002000400007&lng=es&nrm=iso}
}

@article{karch2025probing,
  title={Probing quantum many-body dynamics using subsystem Loschmidt echos},
  author={Karch, Simon and Bandyopadhyay, Souvik and Sun, Zheng-Hang and Impertro, Alexander and Huh, SeungJung and Rodr{\'\i}guez, Irene Prieto and Wienand, Julian F and Ketterle, Wolfgang and Heyl, Markus and Polkovnikov, Anatoli and others},
  journal={arXiv preprint arXiv:2501.16995},
  year={2025}
}

@article{kmvc-88rh,
  title = {Degeneracy beyond the parity-symmetry protection in one-dimensional spinless models: The parity-violating {K}err parametric oscillator},
  author = {Khalouf-Rivera, Jamil and Carvajal, Miguel and P\'erez-Bernal, Francisco},
  journal = {Phys. Rev. A},
  volume = {113},
  issue = {5},
  pages = {052206},
  numpages = {8},
  year = {2026},
  month = {May},
  publisher = {American Physical Society},
  doi = {10.1103/kmvc-88rh},
  url = {https://link.aps.org/doi/10.1103/kmvc-88rh}
}

@article{QuantumOptics,
title = {QuantumOptics.jl: A Julia framework for simulating open quantum systems},
journal = {Comput. Phys. Commun.},
volume = {227},
pages = {109-116},
year = {2018},
issn = {0010-4655},
doi = {https://doi.org/10.1016/j.cpc.2018.02.004},
url = {https://www.sciencedirect.com/science/article/pii/S0010465518300328},
author = {Sebastian Krämer and David Plankensteiner and Laurin Ostermann and Helmut Ritsch}
}

@article{Link2020,
  title = {Dynamical Phase Transitions in Dissipative Quantum Dynamics with Quantum Optical Realization},
  author = {Link, Valentin and Strunz, Walter T.},
  journal = {Phys. Rev. Lett.},
  volume = {125},
  issue = {14},
  pages = {143602},
  numpages = {6},
  year = {2020},
  month = {Sep},
  publisher = {American Physical Society},
  doi = {10.1103/PhysRevLett.125.143602},
  url = {https://link.aps.org/doi/10.1103/PhysRevLett.125.143602}
}

@article{lohr2025dynamical,
  title={Dynamical quantum phase transitions in the {L}ipkin--{M}eshkov--{G}lick model with longitudinal and transverse magnetic fields},
  author={Lohr-Robles, DS and Grether, M and L{\'o}pez-Moreno, E and Hess, PO},
  journal={Eur. Phys. J. B},
  volume={98},
  number={12},
  pages={260},
  year={2025},
  publisher={Springer},
  doi={https://doi.org/10.1140/epjb/s10051-025-01111-0}
}

@article{LopezPena2021,
doi = {10.1088/1402-4896/abd654},
url = {https://dx.doi.org/10.1088/1402-4896/abd654},
year = {2021},
month = {jan},
publisher = {IOP Publishing},
volume = {96},
number = {3},
pages = {035103},
author = {López-Peña, R and Cordero, S and Nahmad-Achar, E and Castaños, O},
title = {Quantum phase diagrams of matter-field Hamiltonians II: Wigner function analysis},
journal = {Phys. Scr.}
}

@article{lv2018quantum,
  title={Quantum simulation of the quantum {R}abi model in a trapped ion},
  author={Lv, Dingshun and An, Shuoming and Liu, Zhenyu and Zhang, Jing-Ning and Pedernales, Julen S and Lamata, Lucas and Solano, Enrique and Kim, Kihwan},
  journal={Phys. Rev. X},
  volume={8},
  number={2},
  pages={021027},
  year={2018},
  publisher={APS}
}

@book{larson2021jaynes,
  title={The Jaynes--Cummings model and its descendants: modern research directions},
  author={Larson, Jonas and Mavrogordatos, Themistoklis},
  year={2021},
  publisher={IoP Publishing},
  url = {https://doi.org/10.1088/978-0-7503-3447-1},
  doi = {10.1088/978-0-7503-3447-1}
}

@article{liu2022experimental,
  title={Experimental demonstration of remotely creating {W}igner negativity via quantum steering},
  author={Liu, Shuheng and Han, Dongmei and Wang, Na and Xiang, Yu and Sun, Fengxiao and Wang, Meihong and Qin, Zhongzhong and Gong, Qihuang and Su, Xiaolong and He, Qiongyi},
  journal={Phys. Rev. Lett. },
  volume={128},
  number={20},
  pages={200401},
  year={2022},
  publisher={APS},
  doi = {10.1103/PhysRevLett.128.200401},
  url = {https://link.aps.org/doi/10.1103/PhysRevLett.128.200401}
}

@article{langford2017experimentally,
  title={Experimentally simulating the dynamics of quantum light and matter at deep-strong coupling},
  author={Langford, Nathan K and Sagastizabal, Ramiro and Kounalakis, Marios and Dickel, Christian and Bruno, Alessandro and Luthi, Florian and Thoen, David J and Endo, Akira and DiCarlo, Leonardo},
  journal={Nat. Commun.},
  volume={8},
  number={1},
  pages={1715},
  year={2017},
  publisher={Nature Publishing Group UK London},
  url={https://doi.org/10.1038/s41467-017-01061-x}
}

@article{Mera2018,
  title = {Dynamical phase transitions at finite temperature from fidelity and interferometric {L}oschmidt echo induced metrics},
  author = {Mera, Bruno and Vlachou, Chrysoula and Paunkovi{\ifmmode \acute{c}\else \'{c}\fi{}}, Nikola and Vieira, V{\'{\i}}tor R. and Viyuela, Oscar},
  journal = {Phys. Rev. B},
  volume = {97},
  issue = {9},
  pages = {094110},
  numpages = {15},
  year = {2018},
  month = {Mar},
  publisher = {American Physical Society},
  doi = {10.1103/PhysRevB.97.094110},
  url = {https://link.aps.org/doi/10.1103/PhysRevB.97.094110}
}

@article{Mirrahimi_2014,
doi = {10.1088/1367-2630/16/4/045014},
url = {https://doi.org/10.1088/1367-2630/16/4/045014},
year = {2014},
month = {apr},
publisher = {IOP Publishing},
volume = {16},
number = {4},
pages = {045014},
author = {Mirrahimi, Mazyar and Leghtas, Zaki and Albert, Victor V and Touzard, Steven and Schoelkopf, Robert J and Jiang, Liang and Devoret, Michel H},
title = {Dynamically protected cat-qubits: a new paradigm for universal quantum computation},
journal = {New Journal of Physics}
}

@article{marti2024kerr,
  title={Quantum squeezing in a nonlinear mechanical oscillator},
  author={Marti, Stefano and von L{\"u}pke, Uwe and Joshi, Om and Yang, Yu and Bild, Marius and Omahen, Andraz and Chu, Yiwen and Fadel, Matteo},
  journal={Nat. Phys},
  pages={1--6},
  year={2024},
  publisher={Nature Publishing Group UK London},
  doi = {https://doi.org/10.1038/s41567-024-02545-6}
}

@book{NielsenChuang2010,
  author    = {Michael A. Nielsen and Isaac L. Chuang},
  title     = {Quantum Computation and Quantum Information},
  publisher = {Cambridge University Press},
  year      = {2010},
  doi       = {10.1017/CBO9780511976667}
}

@article{NaderBEC2025,
author = {Nader, D. J. and Serrano-Ensástiga, E.},
title = {Critical Energies and {W}igner Functions of the Stationary States of the {B}ose {E}instein Condensates in a Double-Well Trap},
journal = {Adv. Quantum Technol.},
volume = {8},
number = {7},
pages = {2400451},
doi = {https://doi.org/10.1002/qute.202400451},
year = {2025}
}

@article{Nader2021PRE,
  title = {Avoided crossings and dynamical tunneling close to excited-state quantum phase transitions},
  author = {Nader, D. J. and Gonz\'alez-Rodr\'{\i}guez, C. A. and Lerma-Hern\'andez, S.},
  journal = {Phys. Rev. E},
  volume = {104},
  issue = {6},
  pages = {064116},
  numpages = {15},
  year = {2021},
  month = {Dec},
  publisher = {American Physical Society},
  doi = {10.1103/PhysRevE.104.064116},
  url = {https://link.aps.org/doi/10.1103/PhysRevE.104.064116}
}

@article{Macek2019,
  title = {Excited-state quantum phase transitions in systems with two degrees of freedom. III. Interacting boson systems},
  author = {Macek, Michal and Str\'ansk\'y, Pavel and Leviatan, Amiram and Cejnar, Pavel},
  journal = {Phys. Rev. C},
  volume = {99},
  issue = {6},
  pages = {064323},
  numpages = {14},
  year = {2019},
  month = {Jun},
  publisher = {American Physical Society},
  doi = {10.1103/PhysRevC.99.064323},
  url = {https://link.aps.org/doi/10.1103/PhysRevC.99.064323}
}

@article{Mari2012,
  title = {Positive {W}igner Functions Render Classical Simulation of Quantum Computation Efficient},
  author = {Mari, A. and Eisert, J.},
  journal = {Phys. Rev. Lett.},
  volume = {109},
  issue = {23},
  pages = {230503},
  numpages = {5},
  year = {2012},
  month = {Dec},
  publisher = {American Physical Society},
  doi = {10.1103/PhysRevLett.109.230503},
  url = {https://link.aps.org/doi/10.1103/PhysRevLett.109.230503}
}

@article{Pan_2010,
doi = {10.1088/0953-4075/43/17/175501},
url = {https://doi.org/10.1088/0953-4075/43/17/175501},
year = {2010},
month = {aug},
publisher = {},
volume = {43},
number = {17},
pages = {175501},
author = {Pan, Feng and Guan, Xin and Wang, Yin and Draayer, J P},
title = {A progressive diagonalization scheme for the {R}abi {H}amiltonian},
journal = {J. Phys. B: At. Mol. Opt. Phys.}
}

@article{PueblaQPT_QRM,
  title = {Quantum Phase Transition and Universal Dynamics in the {R}abi Model},
  author = {Hwang, Myung-Joong and Puebla, Ricardo and Plenio, Martin B.},
  journal = {Phys. Rev. Lett.},
  volume = {115},
  issue = {18},
  pages = {180404},
  numpages = {5},
  year = {2015},
  month = {Oct},
  publisher = {American Physical Society},
  doi = {10.1103/PhysRevLett.115.180404},
  url = {https://link.aps.org/doi/10.1103/PhysRevLett.115.180404}
}

@article{park2022slowing,
  title={Slowing quantum decoherence of oscillators by hybrid processing},
  author={Park, Kimin and Hastrup, Jacob and Neergaard-Nielsen, Jonas Schou and Brask, Jonatan Bohr and Filip, Radim and Andersen, Ulrik L},
  journal={npj Quantum Inf },
  volume={8},
  number={1},
  pages={67},
  year={2022},
  publisher={Nature Publishing Group UK London},
  url = {https://doi.org/10.1038/s41534-022-00577-5}
}

@article{Prado2023,
  title = {Quantum tunneling and level crossings in the squeeze-driven {K}err oscillator},
  author = {Reynoso, Miguel A. Prado and Nader, D. J. and Ch\'avez-Carlos, Jorge and Ordaz-Mendoza, B. E. and Corti\~nas, Rodrigo G. and Batista, Victor S. and Lerma-Hern\'andez, S. and P\'erez-Bernal, Francisco and Santos, Lea F.},
  journal = {Phys. Rev. A},
  volume = {108},
  issue = {3},
  pages = {033709},
  numpages = {12},
  year = {2023},
  month = {Sep},
  publisher = {American Physical Society},
  doi = {10.1103/PhysRevA.108.033709},
  url = {https://link.aps.org/doi/10.1103/PhysRevA.108.033709}
}

@article{Peota2021,
  title = {Determination of Dynamical Quantum Phase Transitions in Strongly Correlated Many-Body Systems Using {L}oschmidt Cumulants},
  author = {Peotta, Sebastiano and Brange, Fredrik and Deger, Aydin and Ojanen, Teemu and Flindt, Christian},
  journal = {Phys. Rev. X},
  volume = {11},
  issue = {4},
  pages = {041018},
  numpages = {15},
  year = {2021},
  month = {Oct},
  publisher = {American Physical Society},
  doi = {10.1103/PhysRevX.11.041018},
  url = {https://link.aps.org/doi/10.1103/PhysRevX.11.041018}
}

@article{Puebla2020,
  title = {Finite-component dynamical quantum phase transitions},
  author = {Puebla, Ricardo},
  journal = {Phys. Rev. B},
  volume = {102},
  issue = {22},
  pages = {220302},
  numpages = {6},
  year = {2020},
  month = {Dec},
  publisher = {American Physical Society},
  doi = {10.1103/PhysRevB.102.220302},
  url = {https://link.aps.org/doi/10.1103/PhysRevB.102.220302}
}

@article{polkovnikov2005universal,
  title={Universal adiabatic dynamics in the vicinity of a quantum critical point},
  author={Polkovnikov, Anatoli},
  journal={Phys. Rev. B},
  volume={72},
  number={16},
  pages={161201},
  year={2005},
  publisher={APS},
  doi = {10.1103/PhysRevB.72.161201},
  url = {https://link.aps.org/doi/10.1103/PhysRevB.72.161201}
}

@article{Puchala2009,
  title = {Bound on trace distance based on superfidelity},
  author = {Pucha\l{}a, Zbigniew and Miszczak, Jaros\l{}aw Adam},
  journal = {Phys. Rev. A},
  volume = {79},
  issue = {2},
  pages = {024302},
  numpages = {4},
  year = {2009},
  month = {Feb},
  publisher = {American Physical Society},
  doi = {10.1103/PhysRevA.79.024302},
  url = {https://link.aps.org/doi/10.1103/PhysRevA.79.024302}
}

@article{POLKOVNIKOV_TCA,
title = {Phase space representation of quantum dynamics},
journal = {Ann. Phys.},
volume = {325},
number = {8},
pages = {1790-1852},
year = {2010},
issn = {0003-4916},
doi = {https://doi.org/10.1016/j.aop.2010.02.006},
url = {https://www.sciencedirect.com/science/article/pii/S0003491610000382},
author = {Anatoli Polkovnikov}
}

@article{Puebla2016,
  title = {Excited-state quantum phase transition in the {R}abi model},
  author = {Puebla, Ricardo and Hwang, Myung-Joong and Plenio, Martin B.},
  journal = {Phys. Rev. A},
  volume = {94},
  issue = {2},
  pages = {023835},
  numpages = {9},
  year = {2016},
  month = {Aug},
  publisher = {American Physical Society},
  doi = {10.1103/PhysRevA.94.023835},
  url = {https://link.aps.org/doi/10.1103/PhysRevA.94.023835}
}

@article{rabi1936process,
  title={On the process of space quantization},
  author={Rabi, II},
  journal={Phys. Rev.},
  volume={49},
  number={4},
  pages={324},
  year={1936},
  publisher={APS},
  doi = {10.1103/PhysRev.49.324},
  url = {https://link.aps.org/doi/10.1103/PhysRev.49.324}
}

@article{rackauckas2017differentialequations,
  title={DifferentialEquations.jl – A Performant and Feature-Rich Ecosystem for
 Solving Differential Equations in Julia},
  author={Christopher Rackauckas and Qing Nie},
  journal={J. Open Res. Softw.},
  year={2017},
  volume={5},
  pages={15},
  url={https://api.semanticscholar.org/CorpusID:64912314}
}

@article{Shore01071993,
author = {Bruce W. Shore and Peter L. Knight},
title = {The {J}aynes-{C}ummings Model},
journal = {J. Mod. Opt.},
volume = {40},
number = {7},
pages = {1195--1238},
year = {1993},
publisher = {Taylor \& Francis},
doi = {10.1080/09500349314551321},


URL = { 
    
        https://doi.org/10.1080/09500349314551321
    
    

},
eprint = { 
    
        https://doi.org/10.1080/09500349314551321
    
    

}

}

@article{SantosESQPT,
  title = {Structure of eigenstates and quench dynamics at an excited-state quantum phase transition},
  author = {Santos, Lea F. and P\'erez-Bernal, Francisco},
  journal = {Phys. Rev. A},
  volume = {92},
  issue = {5},
  pages = {050101},
  numpages = {5},
  year = {2015},
  month = {Nov},
  publisher = {American Physical Society},
  doi = {10.1103/PhysRevA.92.050101},
  url = {https://link.aps.org/doi/10.1103/PhysRevA.92.050101}
}

@article{Sedlmayr2018,
  title = {Fate of dynamical phase transitions at finite temperatures and in open systems},
  author = {Sedlmayr, N. and Fleischhauer, M. and Sirker, J.},
  journal = {Phys. Rev. B},
  volume = {97},
  issue = {4},
  pages = {045147},
  numpages = {8},
  year = {2018},
  month = {Jan},
  publisher = {American Physical Society},
  doi = {10.1103/PhysRevB.97.045147},
  url = {https://link.aps.org/doi/10.1103/PhysRevB.97.045147}
}

@article{sen2005dynamical,
  title={Dynamical phase transitions and temperature-induced quantum correlations in an infinite spin chain},
  author={Sen, Aditi and Sen, Ujjwal and Lewenstein, Maciej},
  journal={Phys. Rev. A.},
  volume={72},
  number={5},
  pages={052319},
  year={2005},
  publisher={APS},
  doi = {10.1103/PhysRevA.72.052319},
  url = {https://link.aps.org/doi/10.1103/PhysRevA.72.052319}
}

@article{smithey1993measurement,
  title={Measurement of the {W}igner distribution and the density matrix of a light mode using optical homodyne tomography: Application to squeezed states and the vacuum},
  author={Smithey, Daniel T and Beck, Mark and Raymer, Michael G and Faridani, Adel},
  journal={Phys. Rev. Lett.},
  volume={70},
  number={9},
  pages={1244},
  year={1993},
  publisher={APS},
  doi = {10.1103/PhysRevLett.70.1244},
  url = {https://link.aps.org/doi/10.1103/PhysRevLett.70.1244}
}

@inbook{Sachdev_2011, place={Cambridge}, title={Introduction}, booktitle={Quantum Phase Transitions},doi={https://doi.org/10.1017/CBO9780511973765 }, publisher={Cambridge University Press}, author={Sachdev, Subir}, year={2011}, pages={1–2}}

@article{catdeath2024,
  title = {Unitary collapse of {S}chr\"odinger's cat state},
  author = {Str\'ansk\'y, Pavel and Cejnar, Pavel and Filip, Radim},
  journal = {Phys. Rev. A},
  volume = {110},
  issue = {3},
  pages = {L030202},
  numpages = {6},
  year = {2024},
  month = {Sep},
  publisher = {American Physical Society},
  doi = {10.1103/PhysRevA.110.L030202},
  url = {https://link.aps.org/doi/10.1103/PhysRevA.110.L030202}
}

@article{stransky2015excited,
title = {Excited-state quantum phase transitions in systems with two degrees of freedom: {II}. Finite-size effects},
journal = {Ann. Phys.},
volume = {356},
pages = {57-82},
year = {2015},
issn = {0003-4916},
doi = {https://doi.org/10.1016/j.aop.2015.02.025},
url = {https://www.sciencedirect.com/science/article/pii/S0003491615000767},
author = {Pavel Stránský and Michal Macek and Amiram Leviatan and Pavel Cejnar}
}

@article{stransky2014excited,
title = {Excited-state quantum phase transitions in systems with two degrees of freedom: Level density, level dynamics, thermal properties},
journal = {Ann. Phys.},
volume = {345},
pages = {73-97},
year = {2014},
issn = {0003-4916},
doi = {https://doi.org/10.1016/j.aop.2014.03.006},
url = {https://www.sciencedirect.com/science/article/pii/S000349161400061X},
author = {Pavel Stránský and Michal Macek and Pavel Cejnar}
}

@article{Stransky2021,
  title = {Stabilization of product states and excited-state quantum phase transitions in a coupled qubit-field system},
  author = {Str\'ansk\'y, Pavel and Cejnar, Pavel and Filip, Radim},
  journal = {Phys. Rev. A},
  volume = {104},
  issue = {5},
  pages = {053722},
  numpages = {14},
  year = {2021},
  month = {Nov},
  publisher = {American Physical Society},
  doi = {10.1103/PhysRevA.104.053722},
  url = {https://link.aps.org/doi/10.1103/PhysRevA.104.053722}
}

@article{TWA_2015,
  title = {Many-Body Quantum Spin Dynamics with {M}onte {C}arlo Trajectories on a Discrete Phase Space},
  author = {Schachenmayer, J. and Pikovski, A. and Rey, A. M.},
  journal = {Phys. Rev. X},
  volume = {5},
  issue = {1},
  pages = {011022},
  numpages = {10},
  year = {2015},
  month = {Feb},
  publisher = {American Physical Society},
  doi = {10.1103/PhysRevX.5.011022},
  url = {https://link.aps.org/doi/10.1103/PhysRevX.5.011022}
}

@article{trenkwalder2016quantum,
  title={Quantum phase transitions with parity-symmetry breaking and hysteresis},
  author={Trenkwalder, A and Spagnolli, G and Semeghini, Giulia and Coop, S and Landini, M and Castilho, P and Pezze, L and Modugno, Giovanni and Inguscio, Massimo and Smerzi, A and others},
  journal={Nature physics},
  volume={12},
  number={9},
  pages={826--829},
  year={2016},
  publisher={Nature Publishing Group UK London},
  doi={https://doi.org/10.1038/nphys3743}
}

@article{todorov2009strong,
  title={Strong Light-Matter Coupling in Subwavelength Metal-Dielectric Microcavities at Terahertz Frequencies},
  author={Todorov, Y and Andrews, Aaron Maxwell and Sagnes, I and Colombelli, R and Klang, Pavel and Strasser, Gottfried and Sirtori, C},
  journal={Phys. Rev. Lett.},
  volume={102},
  number={18},
  pages={186402},
  year={2009},
  publisher={APS},
  doi = {10.1103/PhysRevLett.102.186402},
  url = {https://link.aps.org/doi/10.1103/PhysRevLett.102.186402}
}

@article{UHLMANN1976273,
title = {The transition probability in the state space of algebra},
journal = {Rep. Math. Phys.},
volume = {9},
number = {2},
pages = {273-279},
year = {1976},
issn = {0034-4877},
doi = {https://doi.org/10.1016/0034-4877(76)90060-4},
url = {https://www.sciencedirect.com/science/article/pii/0034487776900604},
author = {A. Uhlmann},
}

@article{vosk2014dynamical,
  title={Dynamical quantum phase transitions in random spin chains},
  author={Vosk, Ronen and Altman, Ehud},
  journal={Phys. Rev. Lett.},
  volume={112},
  number={21},
  pages={217204},
  year={2014},
  publisher={APS},
  doi = {10.1103/PhysRevLett.112.217204},
  url = {https://link.aps.org/doi/10.1103/PhysRevLett.112.217204}
}

@article{Villasenor_2020,
doi = {10.1088/1367-2630/ab8ef8},
url = {https://dx.doi.org/10.1088/1367-2630/ab8ef8},
year = {2020},
month = {jun},
publisher = {IOP Publishing},
volume = {22},
number = {6},
pages = {063036},
author = {D Villaseñor and S Pilatowsky-Cameo and M A Bastarrachea-Magnani and S Lerma-Hernández and L F Santos and J G Hirsch},
title = {Quantum vs classical dynamics in a spin-boson system: manifestations of spectral correlations and scarring},
journal = {New J. Phys.}
}

@article{vojta2003quantum,
  title={Quantum phase transitions},
  author={Vojta, Matthias},
  journal={Rep. Prog. Phys.},
  volume={66},
  number={12},
  pages={2069},
  year={2003},
  publisher={IOP Publishing},
  doi={10.1088/0034-4885/66/12/R01}
}

@article{vanicek2003survival,
  title={Survival probability and local density of states for one-dimensional {H}amiltonian systems},
  author={Vaníček, Jiří and Cohen, Doron},
  journal={J. Phys. A: Math. Gen. },
  volume={36},
  number={36},
  pages={9591--9608},
  year={2003},
  doi={10.1088/0305-4470/36/36/310}
}

@article{Vajna2014,
  title = {Disentangling dynamical phase transitions from equilibrium phase transitions},
  author = {Vajna, Szabolcs and D{\'o}ra, Bal{\'a}zs},
  journal = {Phys. Rev. B},
  volume = {89},
  issue = {16},
  pages = {161105(R)},
  numpages = {5},
  year = {2014},
  month = {Apr},
  publisher = {American Physical Society},
  doi = {10.1103/PhysRevB.89.161105},
  url = {https://link.aps.org/doi/10.1103/PhysRevB.89.161105}
}

@ARTICLE{wignerfunctions,
       author = {{Case}, William B.},
        title = "{Wigner functions and Weyl transforms for pedestrians}",
      journal = { Am. J. Phys.},
         year = 2008,
        month = oct,
       volume = {76},
       number = {10},
        pages = {937-946},
          doi = {10.1119/1.2957889},
       adsurl = {https://ui.adsabs.harvard.edu/abs/2008AmJPh..76..937C}
}

@article{wang2021esqpt,
  title = {Signatures of excited-state quantum phase transitions in quantum many-body systems: Phase space analysis},
  author = {Wang, Qian and P\'erez-Bernal, Francisco},
  journal = {Phys. Rev. E},
  volume = {104},
  issue = {3},
  pages = {034119},
  numpages = {15},
  year = {2021},
  month = {Sep},
  publisher = {American Physical Society},
  doi = {10.1103/PhysRevE.104.034119},
  url = {https://link.aps.org/doi/10.1103/PhysRevE.104.034119}
}

@article{wang2009decoherence,
  title={Decoherence dynamics of complex photon states in a superconducting circuit},
  author={Wang, H and Hofheinz, M and Ansmann, M and Bialczak, RC and Lucero, Erik and Neeley, M and O’Connell, AD and Sank, D and Weides, M and Wenner, J and others},
  journal={Phys. Rev. Lett.},
  volume={103},
  number={20},
  pages={200404},
  year={2009},
  publisher={APS},
  doi = {10.1103/PhysRevLett.103.200404},
  url = {https://link.aps.org/doi/10.1103/PhysRevLett.103.200404}
}

@article{wakui2007photon,
  title={Photon subtracted squeezed states generated with periodically poled KTiOPO4},
  author={Wakui, Kentaro and Takahashi, Hiroki and Furusawa, Akira and Sasaki, Masahide},
  journal={Opt. Express},
  volume={15},
  number={6},
  pages={3568--3574},
  year={2007},
  publisher={Optical Society of America},
  url = {https://opg.optica.org/oe/abstract.cfm?URI=oe-15-6-3568},
  doi = {10.1364/OE.15.003568},
}

@article{wang2016schrodinger,
  title={A {S}chr{\"o}dinger cat living in two boxes},
  author={Wang, Chen and Gao, Yvonne Y and Reinhold, Philip and Heeres, Reinier W and Ofek, Nissim and Chou, Kevin and Axline, Christopher and Reagor, Matthew and Blumoff, Jacob and Sliwa, KM and others},
  journal={Science},
  volume={352},
  number={6289},
  pages={1087--1091},
  year={2016},
  publisher={American Association for the Advancement of Science},
  doi = {10.1126/science.aaf2941}
}

@article{PhysRevA.102.063531,
  title = {Excited-state quantum phase transitions in {K}err nonlinear oscillators},
  author = {Wang, Qing-Wei and Wu, Sha},
  journal = {Phys. Rev. A},
  volume = {102},
  issue = {6},
  pages = {063531},
  numpages = {7},
  year = {2020},
  month = {Dec},
  publisher = {American Physical Society},
  doi = {10.1103/PhysRevA.102.063531},
  url = {https://link.aps.org/doi/10.1103/PhysRevA.102.063531}
}

@article{YU2026170344,
title = {Exact solutions and dynamical phase transitions in the {L}ipkin-{M}eshkov-{G}lick model with dual nonlinear interactions},
journal = {Ann. Phys. },
pages = {170344},
year = {2026},
issn = {0003-4916},
doi = {https://doi.org/10.1016/j.aop.2026.170344},
url = {https://www.sciencedirect.com/science/article/pii/S0003491626000035},
author = {Dongyang Yu}
}

@article{yoshihara2017superconducting,
  title={Superconducting qubit--oscillator circuit beyond the ultrastrong-coupling regime},
  author={Yoshihara, Fumiki and Fuse, Tomoko and Ashhab, Sahel and Kakuyanagi, Kosuke and Saito, Shiro and Semba, Kouichi},
  journal={Nat. Phys.},
  volume={13},
  number={1},
  pages={44--47},
  year={2017},
  publisher={Nature Publishing Group UK London},
  url={https://doi.org/10.1038/nphys3906}
}

@article{ZhengPRL,
  title = {Observation of a Superradiant Phase Transition with Emergent Cat States},
  author = {Zheng, Ri-Hua and Ning, Wen and Chen, Ye-Hong and L\"u, Jia-Hao and Shen, Li-Tuo and Xu, Kai and Zhang, Yu-Ran and Xu, Da and Li, Hekang and Xia, Yan and Wu, Fan and Yang, Zhen-Biao and Miranowicz, Adam and Lambert, Neill and Zheng, Dongning and Fan, Heng and Nori, Franco and Zheng, Shi-Biao},
  journal = {Phys. Rev. Lett.},
  volume = {131},
  issue = {11},
  pages = {113601},
  numpages = {7},
  year = {2023},
  month = {Sep},
  publisher = {American Physical Society},
  doi = {10.1103/PhysRevLett.131.113601},
  url = {https://link.aps.org/doi/10.1103/PhysRevLett.131.113601}
}

@article{xu2024quantum,
  title={Quantum-coherence-assisted dynamical phase transition in the one-dimensional transverse-field {I}sing model},
  author={Xu, Bao-Ming},
  journal={Commun. Theor. Phys.},
  volume={76},
  number={12},
  pages={125104},
  year={2024},
  publisher={IOP Publishing},
  doi = {10.1088/1572-9494/ad724c},
}

@article{PhysRevResearch.6.013001,
  title = {Persisting quantum effects in the anisotropic {R}abi model at thermal equilibrium},
  author = {Xu, He-Guang and Montenegro, V. and Xianlong, Gao and Jin, Jiasen and Neto, G. D. de Moraes},
  journal = {Phys. Rev. Res.},
  volume = {6},
  issue = {1},
  pages = {013001},
  numpages = {13},
  year = {2024},
  month = {Jan},
  publisher = {American Physical Society},
  doi = {10.1103/PhysRevResearch.6.013001},
  url = {https://link.aps.org/doi/10.1103/PhysRevResearch.6.013001}
}

@article{xu2020probing,
  title={Probing dynamical phase transitions with a superconducting quantum simulator},
  author={Xu, Kai and Sun, Zheng-Hang and Liu, Wuxin and Zhang, Yu-Ran and Li, Hekang and Dong, Hang and Ren, Wenhui and Zhang, Pengfei and Nori, Franco and Zheng, Dongning and others},
  journal={Sci. Adv.},
  volume={6},
  number={25},
  pages={eaba4935},
  year={2020},
  publisher={American Association for the Advancement of Science},
  doi = {10.1126/sciadv.aba4935}
}

@article{PhysRevE.74.031123,
  title = {Ground state overlap and quantum phase transitions},
  author = {Zanardi, Paolo and Paunkovi\ifmmode \acute{c}\else \'{c}\fi{}, Nikola},
  journal = {Phys. Rev. E},
  volume = {74},
  issue = {3},
  pages = {031123},
  numpages = {6},
  year = {2006},
  month = {Sep},
  publisher = {American Physical Society},
  doi = {10.1103/PhysRevE.74.031123},
  url = {https://link.aps.org/doi/10.1103/PhysRevE.74.031123}
}

@article{novotny2026tracing,
  title={Tracing complex zeros of the quantum survival amplitude: How the energy distribution controls dynamical phase transitions},
  author={Novotn{\'y}, Jakub and St{\v{r}}ele{\v{c}}ek, Jan and Str{\'a}nsk{\'y}, Pavel and Cejnar, Pavel},
  journal={arXiv preprint arXiv:2605.06849},
  year={2026}
}

@incollection{Heller1991LesHouches,
  author    = {Heller, Eric J.},
  title     = {Wavepacket dynamics and quantum chaology},
  booktitle = {Chaos and Quantum Physics (Les Houches Summer School, Session LII, 1989)},
  editor    = {Giannoni, M.-J. and Voros, A. and Zinn-Justin, J.},
  pages     = {547},
  publisher = {North-Holland},
  address   = {Amsterdam},
  year      = {1991}
}

@article{YangLee,
  title = {Statistical Theory of Equations of State and Phase Transitions. {I}. {T}heory of Condensation},
  author = {Yang, C. N. and Lee, T. D.},
  journal = {Phys. Rev.},
  volume = {87},
  issue = {3},
  pages = {404--409},
  numpages = {0},
  year = {1952},
  month = {Aug},
  publisher = {American Physical Society},
  doi = {10.1103/PhysRev.87.404},
  url = {https://link.aps.org/doi/10.1103/PhysRev.87.404}
}

@article{zhou2021exact,
  title={Exact zeros of the {L}oschmidt echo and quantum speed limit time for the dynamical quantum phase transition in finite-size systems},
  author={Zhou, Bozhen and Zeng, Yumeng and Chen, Shu},
  journal={Phys. Rev. B},
  volume={104},
  number={9},
  pages={094311},
  year={2021},
  publisher={APS},
   url = {https://link.aps.org/doi/10.1103/PhysRevB.104.094311}
}

@article{Zvyagin2016DynamicalQP,
  title={Dynamical quantum phase transitions},
  author={Andrei A. Zvyagin},
  journal={Low Temperature Physics},
  year={2016},
  volume={42},
  pages={971-994},
  url={https://api.semanticscholar.org/CorpusID:119453651}
}

\appendix

\section{}\label{proof}

During the evolution in the QRM, the parity operator
$\hat{\Pi}=(-1)^{\hat{a}^\dagger \hat{a}}\hat{\sigma}_z$ is conserved.
Starting from the initial state $|\psi_0\rangle=|0\rangle\otimes |\downarrow\rangle$,
the evolved state can therefore be written as
\begin{equation}
\label{psit}
|\psi(t)\rangle =
\sum_{n\,\mathrm{even}} a_n(t)\,|n\rangle \otimes |\downarrow\rangle
+
\sum_{m\,\mathrm{odd}} b_m(t)\,|m\rangle \otimes |\uparrow\rangle ,
\end{equation}
where $n$ runs over even integers and $m$ over odd integers. The overlap between the initial and evolved states is then
\begin{equation}
\label{overlap}
|{\mathcal Z}(t)|^2 = |\langle \psi_0|\psi(t)\rangle|^2 = |a_0(t)|^2\,.
\end{equation}

The full density matrix reads
\begin{align}
\label{rhofull}
\hat{\rho}(t) &= |\psi(t)\rangle\langle\psi(t)| \nonumber \\
&= \sum_{n,n'} a_n(t)a_{n'}^*(t)\,|n\rangle\langle n'|\otimes|\downarrow\rangle\langle\downarrow| \\
&\quad + \sum_{m,m'} b_m(t)b_{m'}^*(t)\,|m\rangle\langle m'|\otimes|\uparrow\rangle\langle\uparrow| \nonumber \\
&\quad + \sum_{n,m} a_n(t)b_m^*(t)\,|n\rangle\langle m|\otimes|\downarrow\rangle\langle\uparrow| \\
&\quad + \sum_{m,n} b_m(t)a_n^*(t)\,|m\rangle\langle n|\otimes|\uparrow\rangle\langle\downarrow| .
\nonumber
\end{align}
Tracing out the qubit degrees of freedom yields the reduced bosonic density matrix
\begin{eqnarray}
\label{rhob}
\hat{\rho}_b(t)={\rm Tr}_q\big[\hat{\rho}(t)\big]
&=&\sum_{n,n'} a_n(t)a_{n'}^*(t)\,|n\rangle\langle n'|
\nonumber \\ & & +\sum_{m,m'} b_m(t)b_{m'}^*(t)\,|m\rangle\langle m'| .
\end{eqnarray}

The overlap between the initial bosonic density matrix $\hat{\rho}_b(0)=|0\rangle\langle0|$ and the evolved
bosonic density matrix \eqref{rhob} is
\begin{equation}
\label{tr_reduced}
{\rm Tr}\big[\hat{\rho}_b(0)\hat{\rho}_b(t)\big]
=\langle 0|\hat{\rho}_b(t)|0\rangle
=|a_0(t)|^2.
\end{equation}

Comparing with \eqref{overlap}, we find that the survival probability of the full system $|{\mathcal Z}(t)|^2$ can be expressed as the overlap of the initial and the evolved reduced bosonic density matrices. Thus, we can define a bosonic survival probability as $|{\mathcal Z}_b(t)|^2\equiv {\rm Tr} \big[\hat{\rho}_b(t)\hat{\rho}_b(0)\big]$ which contains all relevant information for the analysis of DQPTs.

\section{}\label{Fisher}

In the context of spin--boson systems, the Jaynes--Cummings (JC) model \cite{jaynes1963comparison,Shore01071993,larson2021jaynes} provides the simplest interaction model. Its Hamiltonian is obtained from the QRM \eqref{HQRM} within the rotating-wave approximation by neglecting the counter-rotating terms proportional to $\hat{a}\hat{\sigma}_-$ and $\hat{a}^{\dagger}\hat{\sigma}_+$, resulting in conservation of the total number of excitations. The JC model also exhibits a quantum phase transition at the critical coupling $\lambda_c=2$ \cite{liu2024quantum}. However, unlike the QRM, whose critical point is associated with a second-order QPT \cite{PueblaQPT_QRM}, the JC model undergoes a first-order QPT characterized by a sequence of level crossings between different excitation sectors. In particular, the initial product state $|0\rangle\otimes|\downarrow\rangle$ remains an exact eigenstate of the JC Hamiltonian for arbitrary coupling strength $\lambda$. Consequently, sudden quenches of $\lambda$ starting from this state result in trivial dynamics, with the wave function acquiring only a global phase. The response to quenches in the coupling strength changes qualitatively in the QRM due to the presence of the counter-rotating terms. The product state $|0\rangle\otimes|\downarrow\rangle$ is an eigenstate of the QRM Hamiltonian only at vanishing coupling $\lambda=0$, and therefore quenches to finite coupling strength $\lambda_f$ generate nontrivial dynamics that can lead to nonequilibrium criticality, as shown in Fig.~\ref{Fig1}.

To confirm the nonanalytic behavior of the bosonic rate function i.e. the nonequilibrium criticality in the QRM,  we analyze the complex-time survival amplitude
${\mathcal Z}(t^{\prime})=\sum_k |\langle\psi_0|E_k(\lambda_f)\rangle|^2 e^{-iE_k(\lambda_f)t^\prime}$
with $t^{\prime}=t+i\gamma$, together with its Fisher zeros ${\mathcal Z}(t^\prime)=0$. The survival amplitude is fully characterized by the spectrum of the final Hamiltonian $E_k(\lambda_f)$ and the spectral weights $|\langle\psi_0|E_k(\lambda_f)\rangle|^2$ \cite{novotny2026tracing}. The number of Fisher zeros contained within a closed contour $C$ can be obtained from the contour integral of the logarithmic derivative of the survival amplitude \cite{complextime}. Since the Fisher zeros ${\mathcal Z}(t^{\prime})$ become poles of ${\mathcal Z}'(t^{\prime})/{\mathcal Z}(t^{\prime})$, the residue theorem yields
\begin{equation}
\label{poles}
N_0=\frac{1}{2\pi i}\oint_C \frac{{\mathcal Z}'(t^{\prime})}{{\mathcal Z}(t^\prime)}\,dt^{\prime}
\end{equation}
 where $N_0$ denotes the number of Fisher zeros enclosed by the contour C. Using sufficiently small contours, one can determine the location of the Fisher zeros in the complex-time plane.

\begin{figure}[t!]
    \centering
    \includegraphics[width=\columnwidth]{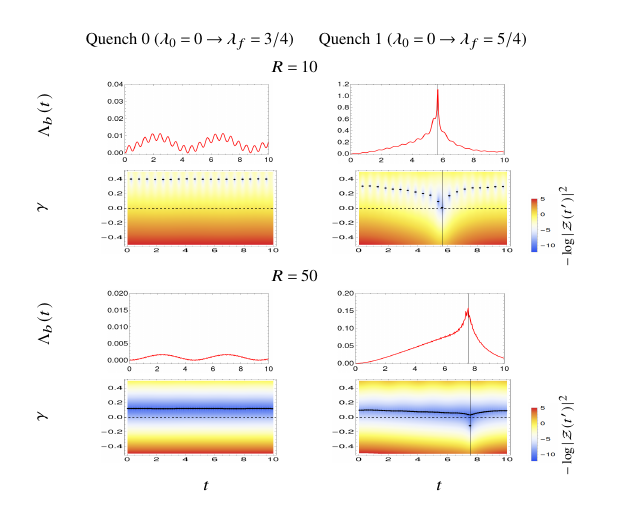}
    \caption{\label{Fisher1} Top row: Bosonic rate function $\Lambda_b(t)$ during representative quenches in the QRM. Bottom row: Negative logarithm of the complex-time survival probability $-\log|{\mathcal Z}(t^\prime)|^2$. The black points are the Fisher zeros ${\mathcal Z}(t^\prime)=0$. The vertical lines indicate the closest Fisher zero to the real axis, which correspondingly manifests in the rate function as a sharp kink. Two representative qubit frequencies are considered: $R=10$ further from the semiclassical limit, and $R=50$, closer to the semiclassical limit.}
\end{figure}

In Fig.~\ref{Fisher1}, we show the distribution of Fisher zeros in the complex-time plane together with the bosonic rate function $\Lambda_b$, to illustrate how the zeros generate peaks. We consider two representative quenches: Quench 0 ($\lambda_0=0 \to \lambda_f=3/4$), with the final coupling in the normal phase, and Quench 1 ($\lambda_0=0 \to \lambda_f=5/4$), with the final coupling in the superradiant phase.
For each, we take two qubit frequencies, $R=10$ and $R=50$, to illustrate the effect of the system size. For the quench within the normal phase ($\lambda_f<\lambda_c$), the Fisher zeros lie only in the positive-$\gamma$ half-plane, while for the quench across the critical point ($\lambda_f>\lambda_c$), zeros also form in the negative-$\gamma$ half-plane.
The resulting structure of zeros crosses the real-time axis, and the zeros lying on or close to the axis give rise to sharp peaks in the rate function.

\begin{figure}[t!]
    \centering
    \includegraphics[width=\columnwidth]{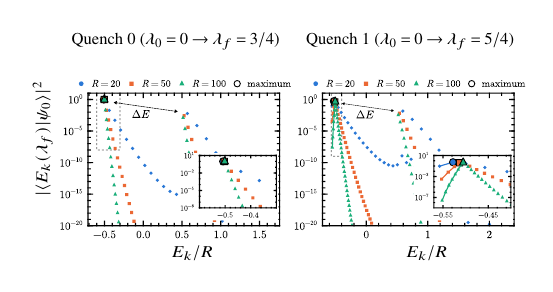}
    \caption{\label{fig:distribution} 
Spectral weights $|\langle E_k(\lambda_f)|\psi_0\rangle|^2$ of the initial state in the eigenbasis of the final Hamiltonian, for qubit frequencies $R=20, 50, 100$.
The insets zoom into the region around the maximum, which is marked by a black outline.
For Quench 0, the maximum lies at the ground state of the final Hamiltonian.
For Quench 1, the maximum is shifted to excited states, which gives rise to a part of the distribution where the envelope increases with energy (emphasized by lines connecting the corresponding points in the inset).
Both distributions are bimodal, with two peaks separated by an energy difference $\Delta E$.
The energy is rescaled as $E_k/R$ to show that the rescaled separation of the peaks is approximately independent of $R$ and $\Delta E$ grows approximately linearly with $R$.}
\end{figure}

This structure of the Fisher zeros in the complex plane can be understood by the theory developed recently in \cite{novotny2026tracing}.
The zeros are approximately organized by the envelope of the energy distribution of the initial state, and the position of its maximum determines whether zeros appear also in the negative-$\gamma$ half-plane.
The part of the envelope on the high-energy side of the maximum, which decreases with energy, generates zeros in the positive-$\gamma$ half-plane.
The part on the low-energy side of the maximum, which increases with energy, generates zeros in the negative-$\gamma$ half-plane.
Fig.~\ref{fig:distribution} shows the energy distributions $|\langle \psi_0|E_k(\lambda_f)\rangle|^2$ of the initial state in the spectrum of the final Hamiltonian, for both quenches and increasing qubit frequency $R$.
For a quench within the normal phase, the most populated state is the ground state of the final Hamiltonian (inset for Quench 0).
The envelope therefore has no part below the maximum, and consequently no zeros form in the negative-$\gamma$ half-plane.
For the quench to the superradiant phase, the maximum moves from the ground state to excited states, and the distribution develops a part at low energies where the envelope increases with energy (inset for Quench 1, where the states forming this part for each qubit frequency $R$ are connected by a line).
The insets of Fig.~\ref{fig:distribution} show that, for Quench 1, the number of states on the low-energy side of the maximum grows with the qubit frequency $R$, which leads to denser formation of Fisher zeros crossing to the negative half-plane, while for Quench 0 the ground state remains the maximum for every $R$.

Before we address the scaling of the Fisher zeros with the system size, let us comment on the prominent dense chains of zeros parallel to the real-time axis in Fig.~\ref{Fisher1}.
The time scale of these chains is set by the bimodal structures of the energy distributions in Fig.~\ref{fig:distribution}.
On a logarithmic scale, the distributions show two peaks separated by an energy difference $\Delta E$, corresponding to the two components of the spin degree of freedom.
The zeros recur approximately with period $\approx 2\pi/\Delta E$ \cite{novotny2026tracing}, and since $\Delta E$ grows approximately linearly with the qubit frequency $R$, the chains become denser in time.
However, the chain parallel to the real time axis give rise only to small, fast oscillations of the rate function, not to the kinks of interest, which come from the shift of the maximum of the energy distribution.

\begin{figure}[htbp]
    \includegraphics[width=0.99\columnwidth]{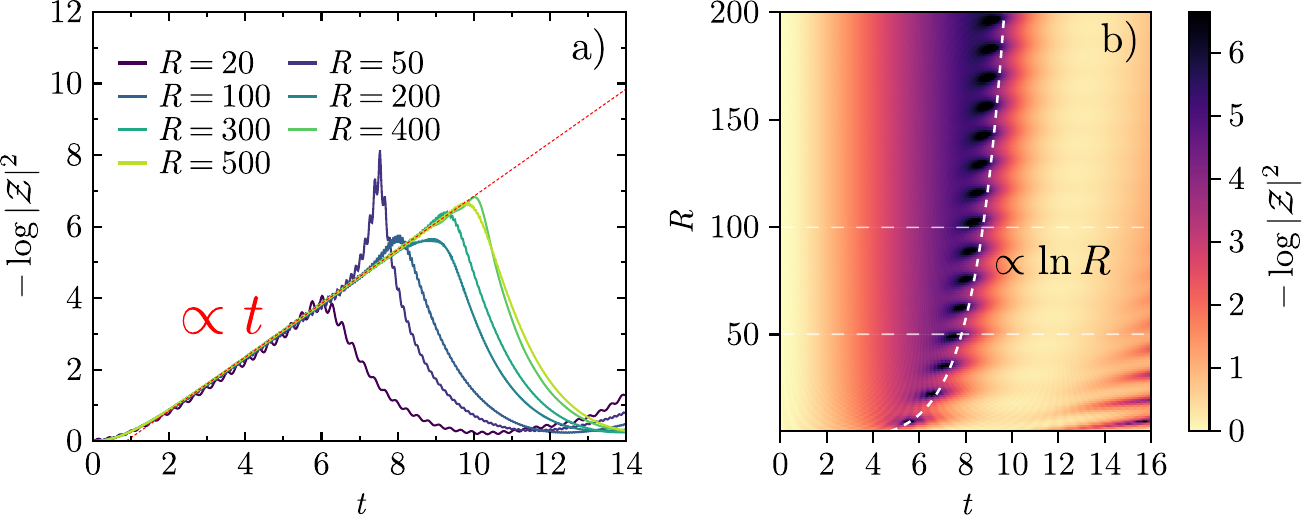}
    \caption{\label{fig:zeros_scaling}
    (a) Logarithm of the survival probability, $-\log|\mathcal Z(t)|^2 = R\,\Lambda_b$, i.e., the rate function without rescaling, for Quench~1 and increasing qubit frequency $R$. The initial growth is linear in time and independent of $R$, as illustrated by the red line.
    (b) Logarithm of the survival probability $-\log|\mathcal Z(t)|^2 = R\,\Lambda_b$ as a function of time and $R$.
    The dark regions corresponds to a Fisher zero on or very close to the real-time axis or equivalently to kinks in the bosonic rate function $\Lambda_b$.
    The white dashed line indicates the $\log R$ shift of the first peak.
    The horizontal dashed lines mark $R=50$ and $R=100$, shown also in panel (a): at $R=50$, a zero lies close to the real-time axis and the kink is sharp, while at $R=100$ it lies farther away, resulting in a smoother peak.
    }
\end{figure}

Fig.~\ref{fig:zeros_scaling}(a) shows the logarithm of the survival probability, $-\log|\mathcal{Z}|^2 = R\,\Lambda_b$, i.e., the rate function without the rescaling by $R$, for Quench~1 and increasing qubit frequency $R$.
The initial peak of the rate function has two important characteristics.
First, its position moves to later times as $\ln R$ with increasing qubit frequency $R$ (see also Fig.~\ref{fig:zeros_scaling}(b)).
Second, before the first peak, the survival probability decays exponentially in time, $|\mathcal{Z}(t)|^2\sim e^{-\Omega t}$, so that $-\log|\mathcal{Z}|^2 \approx \Omega t$, with a rate $\Omega$ independent of $R$.

Fig.~\ref{fig:zeros_scaling}(b) shows the unscaled rate function $-\log|\mathcal Z|^2$ as a function of time and $R$, which reveals how the zeros responsible for the kinks behave during Quench 1 with increasing qubit frequency $R$. It is shown that the zeros reach the real-time axis approximately periodically in $R$, with period $\Delta R \approx 13$.
Each black region corresponds to sharp kinks in the bosonic rate function $\Lambda_b$ and Fisher zeros lying on the real-time axis, as it is for the qubit frequency $R=10$ shown in  Fig.~\ref{Fisher1}. The periodicity of the zeros follows from the envelope picture. As $R$ increases, additional levels join the increasing part of the envelope (connected points in Fig.~\ref{fig:distribution}), and each new level adds one zero to the structure in the negative-$\gamma$ half-plane.
The structure then reorganizes, and one of the Fisher zeros crosses the real-time axis.

\begin{figure}[htbp]
    \includegraphics[width=0.99\columnwidth]{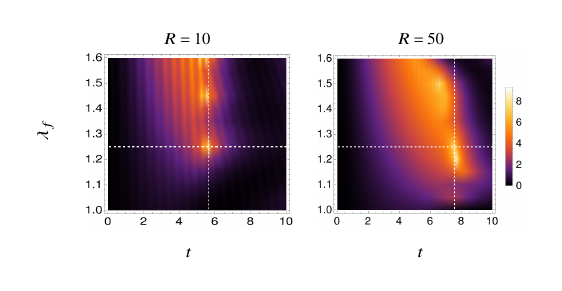}
    \caption{\label{quenches} Negative logarithm of the survival probability
$-{\rm log}|{\mathcal Z}(t)|^2$ as a function of time and final coupling strength $\lambda_f$. The horizontal white dashed lines correspond to the coupling strength $\lambda_f=1.25$ of Quench 1. 
The vertical white dashed lines correspond to the kinks at $t_c=5.65$ and $t_c=7.55$ for $R=10$ and $R=50$, respectively.
    }
\end{figure}

Having confirmed the nonequilibrium criticality from the perspective of Fisher zeros and analyzed the scaling with the qubit frequency $R$,  we analyze the robustness of the kinks against variations in the quench parameters, specifically the final coupling strength $\lambda_f$. For the analysis, we use once again the logarithm of the survival probability $-\log|{\mathcal Z}(t)|^2$. The results are shown in Fig. \ref{quenches} for two representative qubit frequencies $R$. It can be observed that the zeros of the survival probability extends over a wider region of final coupling strength $\lambda_f$ which indicates that the nonequilibrium criticality becomes more robust with increasing qubit frequency $R$.

\section{}\label{surprob}

\begin{figure}[htbp]
    \includegraphics[width=0.99\columnwidth]{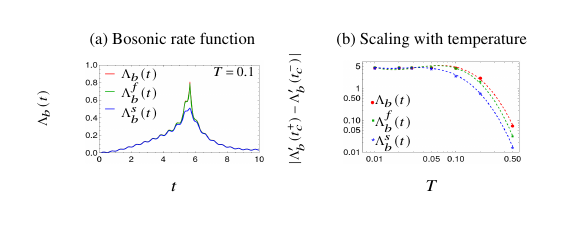}
    \caption{\label{spcomparison} (a) Bosonic rate function $\Lambda_b(t)=-(1/R)\log|{\mathcal Z}_b(t)|^2$.   (b) Scaling with temperature $T$ of the discontinuity in the time derivatives of the bosonic rate function $|\Lambda_b^\prime(t_c^+)-\Lambda_b^\prime(t_c^-)|$.  The dashed lines are fits of the form $f(T)=e^{a_0-a_1\log T -a_2 (\log T)^2}$. For all figures, the initial state is the Gibbs thermal state $\hat{\rho}_{th}(\beta=1/T,\lambda_0=0)$. No superindex indicates bosonic rate function from the overlap $|{\mathcal Z}_b(t)|^2$, super index $f$ indicates bosonic rate function from the fidelity $|{\mathcal Z}_b^{f}(t)|^2$, and superindex $s$ indicates bosonic rate function from the superfidelity $|{\mathcal Z}_b^{s}(t)|^2$.
Both panels correspond to Quench 1 ($\lambda_0 = 0\to \lambda_f=5/4$). The qubit frequency $R=10$ and the loss rate $\kappa=10^{-3}$ are fixed.
    }
\end{figure}

The overlap of the initial and evolved density matrices, namely $|{\mathcal Z}_b(t)|^2 = {\rm Tr}\big[\hat{\rho}_b(t)\hat{\rho}_b(0)\big]$, has been proposed as a natural generalization of the survival probability to the dynamics of mixed states under dissipation \cite{Link2020}. However, it is not unique \cite{Heyl2017DynamicalQP}, and other alternatives are the Uhlmann fidelity \cite{UHLMANN1976273}, $|{\mathcal Z}_b^{f}(t)|^2 = \left({\rm Tr} \sqrt{\sqrt{\hat{\rho}_b(t)}\hat{\rho}_b(0)\sqrt{\hat{\rho}_b(t)}}\right)^2$, and the superfidelity \cite{Puchala2009}, $|{\mathcal Z}_b^{s}(t)|^2 = {\rm Tr}\big[\hat{\rho}_b(t)\hat{\rho}_b(0)\big]+\sqrt{1-{\rm Tr}\hat{\rho}_b^2(0)}\sqrt{1-{\rm Tr}\hat{\rho}_b^2(t)}$. All three quantities correctly reduce to the survival probability for pure states, $|{\mathcal Z}(t)|^2 = |\langle \psi_0|\psi(t)\rangle|^2 = \left|\sum_k |\langle\psi_0|E_k(\lambda_f)\rangle|^2 e^{-iE_k(\lambda_f)t}\right|^2$. Furthermore, if one of the states is pure, namely the initial state $\hat{\rho}_b(0)=|\psi_0\rangle\langle\psi_0|$, all three alternatives reduce to the same value $\langle \psi_0|\hat{\rho}_b(t)|\psi_0\rangle$. For mixed initial states, the superfidelity provides an upper bound on the Uhlmann fidelity \cite{Puchala2009}, while the Uhlmann fidelity, in turn, provides an upper bound on the overlap \cite{NielsenChuang2010}. These relations imply the hierarchy $|{\mathcal Z}_b(t)|^2 \leq |{\mathcal Z}_b^{f}(t)|^2 \leq |{\mathcal Z}_b^{s}(t)|^2$.
Importantly, the overlap $|{\mathcal Z}_b(t)|^2$ vanishes whenever the initial and evolved states become orthogonal, whereas the superfidelity $|{\mathcal Z}_b^{s}(t)|^2$ only vanishes if the states are orthogonal and the initial state is pure. As a consequence, the nonanalytic features (kinks) in the associated rate function $\Lambda_b^{s}(t)=-(1/R)\log|{\mathcal Z}_b^{s}(t)|^2$ become increasingly fragile with growing thermal occupation. By contrast, since the overlap is the smallest among the three measures, it is more susceptible to cancellation. For this reason, it provides a more reliable indicator of sharp nonanalyticities in the rate function $\Lambda_b(t)=-(1/R)\log|{\mathcal Z}_b(t)|^2$, which are characteristic of DQPTs.
This behavior is illustrated in Fig.~\ref{spcomparison}, where numerical results for initial Gibbs states show that the nonanalytic features, specifically, discontinuities in the derivatives of the bosonic rate function, remain more pronounced when using the overlap as the survival probability. In particular, these features scale more slowly with increasing temperature compared to those obtained from the superfidelity, revealing the overlap as the more robust indicator of nonequilibrium criticality during mixed-state dynamics.

\section{}
\label{correspondence}

The semiclassical Hamiltonian \eqref{Hsemiclassical} can be quantized by promoting the momentum to the operator $\hat{p} = -\frac{i}{R}\frac{d}{dx}$, leading to the effective Hamiltonian
\begin{equation}
\label{Hqsemiclassical}
\hat{h}_{eff} = -\frac{1}{2R^2}\frac{d^2}{dx^2} + \frac{1}{2}x^2 - \frac{1}{2}\sqrt{1 + 2\lambda^2x^2}.
\end{equation}
We then evolve the initial state $|\psi_0\rangle = |0\rangle$ under \eqref{Hqsemiclassical} as an approximation to the dynamics of the QRM. This construction corresponds to a requantization of the semiclassical lower-energy branch, in which the qubit acts as an effective ancilla that induces a nonlinear potential in the bosonic mode. As a result, the dynamics is described by a single effective phase space degree of freedom and the state remains pure. This approximation is expected to be valid in the large-$R$ regime and for low-energy dynamics, where the adiabatic elimination of the qubit is justified. However, it does not capture qubit–boson entanglement or the full quantum dynamics at finite $R$, and should therefore be regarded as an effective description of the bosonic sector. The corresponding dynamics under \eqref{Hqsemiclassical} can be obtained numerically using a finite-difference method \cite{NaderBEC2025}.

Using this effective description, we can explain the two features of the first peak in the bosonic rate function identified in Appendix~\ref{Fisher} (Fig.~\ref{fig:zeros_scaling}(a)): the initial exponential decay with an $R$-independent rate, and the $\ln R$ shift of the peak.
The initial state $|\psi_0\rangle=|0\rangle$ corresponds to a Gaussian wave packet centered at the origin of phase space, $x=p=0$.
After the quench to the superradiant phase, the origin becomes an unstable fixed point separating the two wells of the potential.
Expanding the square root in \eqref{Hqsemiclassical} for small $x$ gives, up to a constant, an inverted harmonic oscillator, $\hat h_{eff} \approx \frac{1}{2}\hat p^2 - \frac{1}{2}\Omega^2 x^2$ with $\Omega=\sqrt{\lambda^2-1}$.
Under this Hamiltonian, the wave packet is stretched along the unstable direction as $e^{\Omega t}$, and its survival probability, decays exponentially fast, $|\mathcal Z(t)|^2\sim e^{-\Omega t}$~\cite{Heller1991LesHouches}, independently of the qubit frequency $R$.
For Quench~1, $\Omega=3/4$, in agreement with the slope of the initial linear growth $-\log|\mathcal Z|^2\approx\Omega t$ in Fig.~\ref{fig:zeros_scaling} (a).
This description holds only while the wave packet remains in the region where the quadratic approximation is valid.
This validity breaks down on the order of the Ehrenfest time (following from the exponential stretching of the wave packet near the hyperbolic point), which scales with the effective system size as $\tau_E \propto \ln \hbar_{\rm eff}^{-1} = \ln R$.
The first peak of the rate function, generated by zeros close to the real-time axis, marks this breakdown.
The peak, together with the zeros, shifts to later times as $\ln R$ (Fig.~\ref{fig:zeros_scaling}(b)).
In the limit $R\to\infty$, they move to infinite times.

In addition, we consider a semiclassical approximation based on the \emph{truncated Wigner approximation} (TWA). This method provides a phase space description of the dynamics by truncating higher-order derivatives in the quantum Liouville equation, thereby neglecting the formation of Wigner negativity \cite{TWA_2015,POLKOVNIKOV_TCA,TWA}. The TWA relies on an ensemble of $\mathcal{N}$ stochastic trajectories sampled from the initial Wigner function, which, for the Gaussian state \eqref{W0}, can be treated as a probability distribution. Each trajectory evolves according to the classical Hamilton equations generated by $h_{\mathrm{eff}}$ \eqref{Hsemiclassical}. 

At time $t$, a semiclassical approximation to the phase space distribution is reconstructed from the ensemble of trajectories by forming a histogram over cells of area $\Delta x\,\Delta p$, yielding
\begin{equation}
\label{wignerTWA}
W(x_m,p_n)=\frac{{\mathcal N}_{mn}}{{\mathcal N}\,\Delta x\,\Delta p},
\end{equation}
where ${\mathcal N}_{mn}$ denotes the number of trajectories in the cell centered at $(x_m,p_n)$. By construction, this distribution is strictly positive and therefore cannot reproduce Wigner negativity. Within the TWA approach, the bosonic survival probability can be approximated by evaluating the initial Wigner function at the evolved phase space points \cite{Villasenor_2020,Nader2021PRE}
\begin{equation}
\label{survivalamplitude_TWA}
|{\mathcal Z}_b(t)|^2\approx\frac{2\pi}{R}\frac{1}{\mathcal N}\sum_{i=1}^{\mathcal N} W_0\big(x_i(t),p_i(t)\big).
\end{equation}
Contrary to the exact survival probability, the survival probability from the TWA can not vanish at finite qubit frequency $R$. The corresponding TWA rate function is then obtained by substituting \eqref{survivalamplitude_TWA} into \eqref{ratefunction}. The extension to complex-time in the TWA approach is not as straightforward as it is in the quantum case.

\begin{figure*}[t!]
    \centering
    \includegraphics[width=\textwidth]{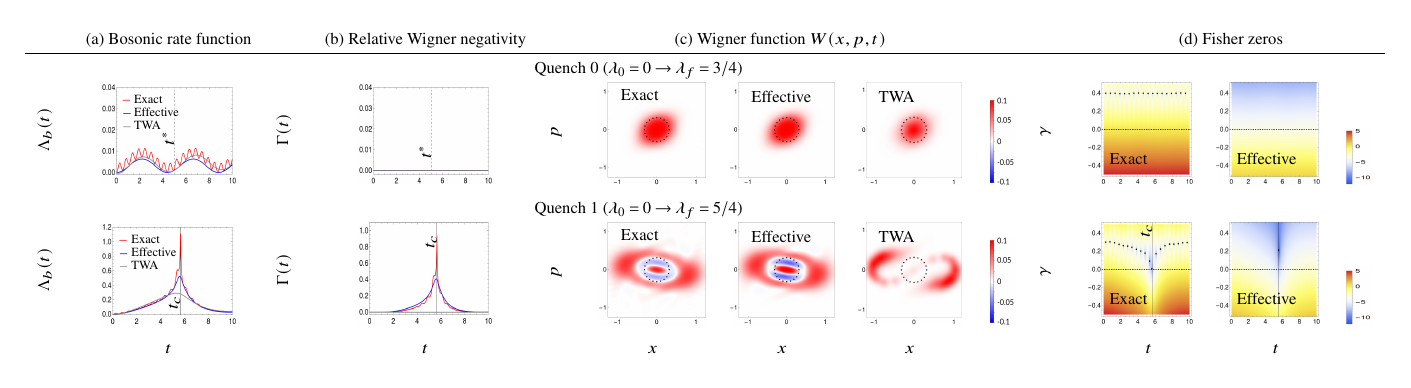}
    \caption{\label{correspondence1} (a) Bosonic rate function $\Lambda_b(t)$. (b) Relative Wigner negativity $\Gamma(t)$. (c) Wigner function at the specific time $t$. (d) Negative logarithm of the survival probability $-\log|{\mathcal Z}(t^\prime)|^2$ with complex argument $t^\prime=t+i\gamma$. The black points are the Fisher zeros ${\mathcal Z}(t^\prime)=0$. The solid vertical lines indicate critical time $t=t_c=5.65$ and the dashed line indicates arbitrary time $t=t^*=5$ for the comparison of the Wigner function. The panels include the exact unitary dynamics of the QRM obtained from the quantum Hamiltonian \eqref{HQRM}, the unitary dynamics generated by the effective Hamiltonian \eqref{Hqsemiclassical}, and the TWA dynamics. For all figures, the qubit frequency is finite $R=10$.}
\end{figure*}

Fig.~\ref{correspondence1} compares the exact quantum dynamics with the two approximations described above. The comparison includes the bosonic rate function, the relative Wigner negativity, and the Wigner function at representative times for two quenches. During Quench 0 with final coupling $\lambda_f=3/4<\lambda_c$ in the normal phase, the dynamics remains close to Gaussian, and all three approaches are in good agreement beyond small oscillations in the rate function. The effective Hamiltonian also reproduces the low-energy spectrum $E_k$ of the QRM rather accurately. Its spectral weights $|\langle\psi_0|E_k\rangle|^2$ capture well whether the maximum of the energy distribution shifts from the ground state to excited states along the quench, which defines if Fisher zeros cross to the negative-$\gamma$ half-plane (see Appendix~\ref{Fisher}). In fact, the effective model essentially reproduces the insets of Fig.~\ref{fig:distribution}.
However, the effective model does not reproduce the full energy distribution as the qubit degree of freedom is neglected. By considering only the low-energy component of the qubit, we discard the high-energy component of the initial state, which is responsible for the second peak in the energy distribution in Fig.~\ref{fig:distribution}. The effective model therefore lacks the bimodal energy distribution and, with it, the chains of Fisher zeros parallel to the real-time axis that produce the oscillations in the bosonic rate function. In any case, the relative negativity is negligible and both the effective Hamiltonian and the TWA correctly reproduce the bosonic Wigner function. 
During Quench 1 with final coupling $\lambda_f=5/4>\lambda_c$ in the superradiant phase, the dynamics becomes strongly non-Gaussian. The effective Hamiltonian \eqref{Hqsemiclassical} partially captures the kink structure in the bosonic rate function and relative Wigner negativity. The sharpness of the kink is however less pronounced under the effective Hamiltonian.
In contrast, the TWA correctly describes the bosonic rate function except near the critical time where the kink emerges. Therefore the TWA fails completely to reproduce the nonequilibrium criticality as the bosonic rate function remains smooth. This is because the TWA cannot reproduce Wigner negativity and its survival probability does not allow the abrupt cancellation necessary for the kink in the rate function.

\begin{figure*}[t!]
    \centering
    \includegraphics[width=0.85\textwidth]{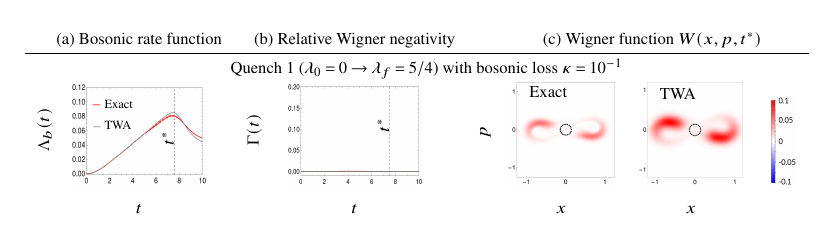}
    \caption{\label{correspondence2} (a) Bosonic rate function $\Lambda_b(t)$, (b) Relative Wigner negativity $\Gamma(t)$ and (c) Wigner function at the specific time $t=t^*=7.55$. The solid vertical lines indicate former critical time $t^*=7.55$ from the unitary dynamics of Fig.~\ref{Fisher1}, for the comparison of the Wigner functions. The panels include the exact open dynamics of the QRM and the TWA dynamics. The bosonic loss rate is $\kappa=10^{-1}$ and the qubit frequency $R=50$.}
\end{figure*}

When dissipation is introduced, however, Wigner negativity is progressively suppressed in the exact model, and the dynamics becomes effectively classical. Increasing the qubit frequency and with strong dissipation, the TWA is expected to progressively recover its validity. To model this situation, we incorporate the bosonic loss into the TWA through a stochastic phase space description based on Langevin equations \cite{gardiner2004quantum,corney2015non}:
\begin{eqnarray}
\label{Langevineq}
\dot{x}&=&\frac{\partial h_{\mathrm{eff}}}{\partial p}-\frac{\kappa}{2}x-\sqrt{\frac{\kappa}{2R}}\,\xi_x(t),\nonumber\\
\dot{p}&=&-\frac{\partial h_{\mathrm{eff}}}{\partial x}-\frac{\kappa}{2}p-\sqrt{\frac{\kappa}{2R}}\,\xi_p(t),
\end{eqnarray}
where $h_{\mathrm{eff}}$ is the semiclassical Hamiltonian~\eqref{Hsemiclassical}. The stochastic terms $\xi_x(t)$ and $\xi_p(t)$ are independent Gaussian white-noise processes with zero mean, $\langle \xi_i(t) \rangle = 0$, and correlations $\langle \xi_i(t)\xi_j(t') \rangle = \delta_{ij}\delta(t-t')$. These noise terms incorporate the effect of dissipation and account for residual fluctuations.
The correspondence is illustrated in Fig.~\ref{correspondence2}, where increasing bosonic loss leads to a progressive smoothing of the Wigner function and a reduction of relative Wigner negativity, while the TWA provides an increasingly accurate description of the dynamics. However, this recovery occurs precisely in the regime where the nonequilibrium criticality disappears: as shown in Fig.~\ref{correspondence2} panel (a), the kinks in the bosonic rate function are smoothed out by dissipation. The suppression of Wigner negativity prevents cancellation between positive and negative phase space contributions necessary for the kink in the relative Wigner negativity. This provides further evidence that the nonequilibrium criticality is intrinsically linked to quantum non-Gaussian effects that cannot be captured within semiclassical descriptions that neglect Wigner negativity.

\section{}\label{Depht}

\begin{figure}[htbp]
    \includegraphics[width=0.99\columnwidth]{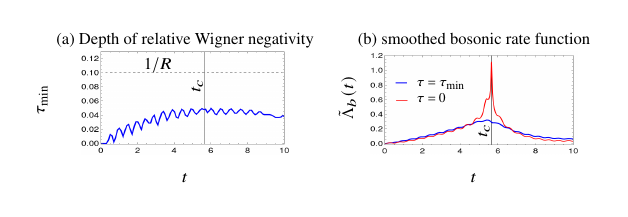}
    \caption{\label{depthfig} (a) Depth $\tau_{\min}$ of the relative Wigner negativity during Quench 1 as a function of time. The black dashed line indicates the maximal depth $1/R$. (b) Smoothed bosonic rate function $\tilde{\Lambda}_b(t)$ during Quench 1 $(\lambda_0=0\to \lambda_f=5/4)$ obtained using the depth $\tau=\tau_{\min}$ (blue line) of the relative Wigner negativity. The bosonic rate function $\tau=0$ (red line) is also shown for comparison. In both panels, the qubit frequency is fixed $R=10$ and the dynamics is unitary $\kappa=0$. The vertical line represents the critical time $t_c=5.65$. 
    }
\end{figure}

The depth of the relative Wigner negativity can be defined in analogy with the depth of nonclassicality of quantum states~\cite{Lee1991}. To this end, we consider a Gaussian smoothing (convolution) of the Wigner function,
\begin{equation}
\label{tranformationW}
\tilde{W}_\tau(x,p,t)=\frac{1}{\pi\tau}\int dx^\prime dp^\prime
e^{-\frac{1}{\tau}\big((x-x^\prime)^2+(p-p^\prime)^2\big)}
W(x^\prime,p^\prime,t),
\end{equation}
where the parameter $ \tau \in [0,1/R] $ controls the amount of smoothing.  In the limit $\tau\to0^+$, the original Wigner function is recovered as $\lim_{\tau\to0^+}\frac{1}{\pi\tau}e^{-\frac{1}{\tau}\big[(x-x^\prime)^2+(p-p^\prime)^2\big]}=\delta(x-x^\prime)\delta(p-p^\prime)$. For $ \tau = 1/R $ the transformation yields the Husimi Q-function, which is positive everywhere~\cite{adesso2014continuous}. The depth of the relative negativity is then defined as the minimal value $ \tau_{\min}$ such that the relative negativity \eqref{Gamma} becomes negligible $\Gamma(t)\lesssim10^{-6}$. In Fig.~\ref{depthfig} panel (a), we show the time dependence of the depth during Quench 1. The depth reaches its maximum at the critical time $t = 5.65$, suggesting that the nonclassicality responsible for the kink in the bosonic rate function is most robust at this point. To observe the effect of the Gaussian smoothing in the bosonic rate function, we define the smoothed survival probability as
\begin{equation}
\label{smooothedSP}
|\tilde{{\mathcal Z}}_b(t)|^2 = \frac{2\pi}{R} \int dx dp W(x,p,0)\tilde{W}_\tau(x,p,t)\,,
\end{equation}
from which the smoothed bosonic rate function  $\tilde{\Lambda}_b(t)=-\frac{1}{R}\log|\tilde{\mathcal Z}_b(t)|^2$ is defined. In Fig.~\ref{depthfig} panel (b), we compare the smoothed and original bosonic rate functions for Quench 1. In the smoothed bosonic rate function, the kink disappears, confirming that the kink is intrinsically linked to the quantum non-Gaussian effects washed out by the Gaussian smoothing.

\section{}\label{oppositequench}

\begin{figure}[t!]
    \centering
    \includegraphics[width=\columnwidth]{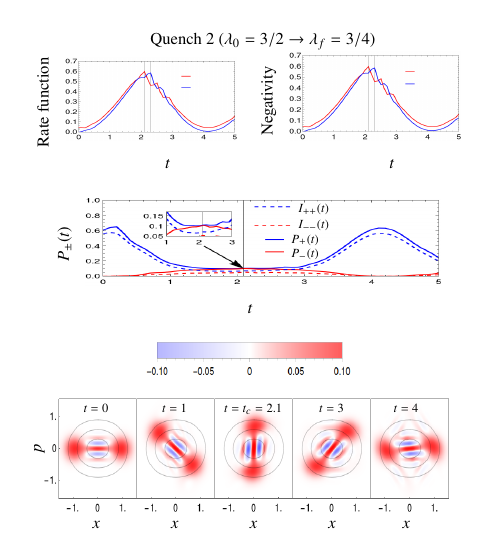}
    \caption{\label{Fig9}
Top left: Rate function of the total spin-boson system $\Lambda(t)=-(1/R)\log|{\mathcal Z}(t)|^2$ and bosonic rate function $\Lambda_b(t)=-(1/R)\log|{\mathcal Z}_b(t)|^2$.  Top right: Wigner negativity indicators 
including global negativity volume $\delta(t)$, global deepest negative value $W_{min}(t)$ and relative negativity $\Gamma(t)$. One representative quench is shown: Quench 2 $(\lambda_0=3/2\to \lambda_f=3/4)$, with initial coupling strength in the superradiant phase and final coupling strength in the normal phase.  Middle row: Phase space contributions $P_\pm(t)$ to the survival probability as a function of time. The vertical lines indicate the kinks at $t_c=2.1$.  Bottom row: Unitary evolution of the Wigner function. The gray lines denote contours of constant energy $h_{\mathrm{eff}}$.  For all figures, the qubit frequency is finite $R=10$.}
\end{figure}

For initial ground states in the superradiant phase $\lambda_0>\lambda_c$, the bosonic mode and the qubit are generally entangled \cite{catrabi}, such that the initial wave function cannot be expressed as a product state. These states are significantly more complicated to prepare experimentally than the initial Gaussian states, but they can be approached starting from the product state 
$|0\rangle\otimes|\downarrow\rangle$
by slowly increasing the coupling strength $\lambda$, obtaining reasonably high fidelities with respect to the ideal ground state sufficiently far from the semiclassical limit \cite{cai2021observation,lv2018quantum,ZhengPRL}. In these cases, the survival probability of the entire spin-boson system  does not reduce to the bosonic survival probability $|{\mathcal Z}_b(t)|^2\neq|{\mathcal Z}(t)|^2$. As a consequence, the rate function of the entire spin-boson system and the bosonic rate function are not identical $\Lambda(t)\neq \Lambda_b(t)$.

An example is shown in Fig.~\ref{Fig9} for Quench 2
($\lambda_0=3/2\to\lambda_f=3/4$). The kink in the bosonic rate
function is only slightly shifted with respect to that of the full spin--boson
rate function. In this case, the initial state already exhibits Wigner negativity, such that both the negativity volume $\delta(t)$ and the deepest negative value $W_{\min}(t)$ are nonzero. Since the initial state is delocalized in phase space, we consider global rather than local negativity quantifiers. Nevertheless, both the negativity volume and the
deepest negative value remain nearly constant throughout
the unitary postquench  and therefore do not provide any signature of the
nonequilibrium criticality. 
In contrast, the relative Wigner negativity $\Gamma(t)$ starts
from zero and develops a pronounced kink that coincides with the kink of
the bosonic rate function. This shows that, also for an initially
non-Gaussian state, the kink in the bosonic rate function can be
attributed to quantum non-Gaussian effects that generate an almost
complete cancellation between positive and negative phase-space
contributions. The unitary evolution of the Wigner function provides a direct
phase-space picture of this mechanism. The state rotates in phase space and the
interference fringes change the orientation facilitating the abrupt cancellation.
This behavior is also reflected in the separate
contributions to the survival probability. The positive contribution
$P_+(t)$ initially differs from unity, reflecting the mixed character
of the reduced bosonic initial state due to qubit--oscillator
entanglement, and subsequently decreases during the evolution. The negative contribution
$P_-(t)$ starts from zero and increases as the interference fringes rotate in phase space.
The phase space contributions $P_-(t)$ and $P_+(t)$ approach each other and remain close over an extended time interval. However, only at the critical time $t_c=2.1$ do they match, causing the bosonic survival probability to vanish and generating the kink in both the bosonic rate function and the relative Wigner negativity. In this quench, the negative phase space contribution $P_-(t) = I_{+-}(t) + I_{-+}(t)$ receives a contribution from the quantum non-Gaussian features developed during the dynamics $I_{+-}(t)$ but also a significant fraction from the initially negative regions that become positive during the dynamics $I_{-+}(t)$. This is consistent with the observed postquench rotation of the interference fringes.

\begin{figure*}[t!]
    \centering
    \includegraphics[width=\textwidth]{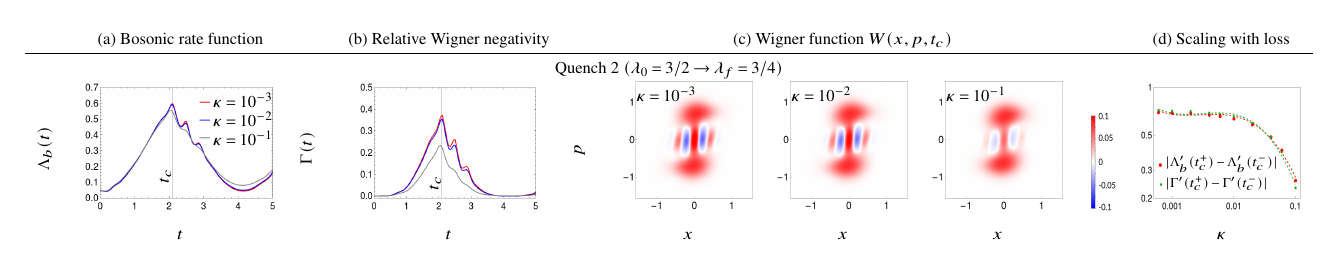}
    \caption{\label{Fig13} (a) Bosonic rate function $\Lambda_b(t)$ as a function of the time $t$. (b) Relative Wigner negativity $\Gamma(t)$. The vertical dashed line marks the critical time. (c) Bosonic Wigner function $W(x,p,t_c)$ at the critical time $t_c=2.1$. (d) Scaling with bosonic loss of the discontinuity in the time derivatives of the bosonic rate function $|\Lambda_b^\prime(t_c^+)-\Lambda_b^\prime(t_c^-)|$ and relative Wigner negativity $|\Gamma^\prime(t_c^+)-\Gamma^\prime(t_c^-)|$. The dashed lines are fits of the form $f(\kappa)=e^{-a-b\log \kappa -c (\log \kappa)^2}$.
For all figures the initial state is a thermal state 
$\hat{\rho}_{th}(\beta=100,\lambda_0=3/2)$ and  the qubit frequency is finite $R=10$.}
\end{figure*}

In Fig. \ref{Fig13}, we investigate the robustness of the nonequilibrium critical behavior of Quench 2 against weak thermal occupation and bosonic loss. For small loss, the characteristic kink in the bosonic rate function $\Lambda_b(t)$ persists together with the corresponding kink in the relative Wigner negativity $\Gamma(t)$, whereas increasing dissipation progressively smooths both features. The Wigner function evaluated at the critical time provides a direct phase-space picture of this degradation, showing that stronger loss suppresses the negative interference fringes responsible for the cancellation between positive and negative contributions to the bosonic survival probability, leading to a smooth bosonic rate function and relative Wigner negativity . The scaling of the discontinuities in the time derivatives of $\Lambda_b(t)$ and $\Gamma(t)$ with the loss rate further confirms their closely related behavior.  Although the kink is less sharp, the scaling is significantly slower for Quench 2 than for Quench 1, suggesting that the characteristic kinks can be more robust against dissipation when the initial state lies in the superradiant phase $\lambda_0>\lambda_c$, where the initial Wigner function already contains negative regions. This is because the negativity of the initial state is not degraded by the loss.

\begin{figure}[H]
    \centering
    \includegraphics[width=0.95\columnwidth]{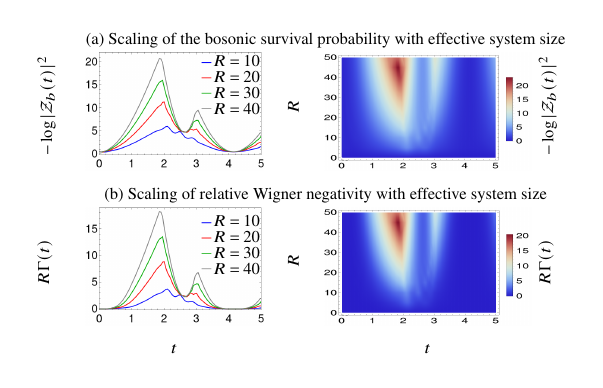}
    \caption{\label{Fig14} (a) Logarithm of the bosonic survival probability $-\log|{\mathcal Z}_b(t)|^2=R\Lambda_b(t)$ without rescaling, with initial thermal occupation $\hat{\rho}_{th}(\beta=100,\lambda_0=3/2)$ and under loss $\kappa=10^{-3}$, for Quench 2 $(\lambda_0=3/2\to\lambda_f=3/4)$. On the left column, we show the initial linear growth with time. On the right column, the logarithm of the survival probability is shown as a function of time $t$ and qubit frequency $R$. The intense red region corresponds to the zero that produces the kink in the bosonic rate function $\Lambda_b(t)$. (b) Same for the relative Wigner negativity $R\Gamma(t)$ without rescaling. }
\end{figure}

Contrary to Quench 1 in the main text, where the kinks in the bosonic rate function $\Lambda_b(t)=-(1/R)\log|{\mathcal Z}_b(t)|^2$ and relative Wigner negativity $\Gamma(t)$ appear and reappear periodically with increasing qubit frequency $R$, in Quench 2 the kinks become increasingly sharp as the qubit frequency $R$ increases, as illustrated in Fig.~\ref{Fig14} under small loss and low thermal occupation.

\end{document}